\documentclass[amsmath,amssymb,twocolumn,superscriptaddress,aps,prb]{revtex4-1}
\pdfoutput=1
\usepackage{silence}
\newlength{\myfigwidth}
\newlength{\myownfigwidth}
\usepackage{amsfonts} 
\usepackage{amsmath}
\usepackage{amssymb}
\usepackage{cancel} 
\usepackage{bm}
\usepackage[usenames, dvipsnames]{color} 
\usepackage{dcolumn}
\usepackage{epic} 
\usepackage{epsfig}
\usepackage{eucal} 
\usepackage{graphicx}
\usepackage[breaklinks,colorlinks = true,linkcolor = red,urlcolor=blue,citecolor=red]{hyperref}
\usepackage{subfigure}
\usepackage{wrapfig} 
\usepackage{xy} 
\usepackage{multirow}
\usepackage{tabularx}

\newcommand{\bcen}{\begin{center}}
\newcommand{\ecen}{\end{center}}
\newcommand{\btab}{\begin{tabular}}
\newcommand{\etab}{\end{tabular}}
\newcommand{\bdes}{\begin{description}}
\newcommand{\edes}{\end{description}}

\newcommand{\beq}{\begin{equation}}
\newcommand{\eeq}{\end{equation}}
\newcommand{\bea}{\begin{eqnarray}}
\newcommand{\eea}{\end{eqnarray}}

\newcommand{\bary}{\begin{array}}
\newcommand{\eary}{\end{array}}
\newcommand{\benum}{\begin{enumerate}}
\newcommand{\eenum}{\end{enumerate}}
\newcommand{\bitem}{\begin{itemize}}
\newcommand{\eitem}{\end{itemize}}

\newcommand{\eqn}[1] {eqn.~(\ref{#1})}

\makeatletter

\newcommand{\Rmnum}[1]{\expandafter\@slowromancap\romannumeral #1@}
\makeatother

\newcommand{\mylabel}[1]{\label{#1}} 

\begin{document}
\title{Interplay of uniform U(1) quantum spin liquid and magnetic phases in rare earth pyrochlore magnets : a fermionic parton approach}


\author{Sambuddha Sanyal}
\email{sambuddhasanyal@gmail.com }
\affiliation{Department of Chemistry, Columbia University, 3000 Broadway, New York 10027, USA}
\author{Kusum Dhochak}
\affiliation{Department of Physics, Indian Institute of Science Education and Research, Bhopal 462066, India}
\author{Subhro Bhattacharjee}
\affiliation{International Centre for Theoretical Sciences, Tata Institute of Fundamental Research,  Bengaluru 560089, India}

\begin{abstract}
We study the uniform time reversal invariant $U(1)$ quantum spin liquid (QSL) with low energy fermionic quasi-particles for rare earth pyrochlore magnets and explore its magnetic instability employing an augmented fermionic parton mean field theory approach. Self consistent calculations stabilise an uniform $U(1)$ QSL with both gapped and gapless parton excitations as well as  fractionalised magnetically ordered phases in an experimentally relevant part of the phase diagram near the classical phase boundaries of the magnetically ordered phases. The gapped QSL has a band-structure with a non-zero $Z_2$ topological invariant. The fractionalised magnetic ordered phases bears signature of both QSL through fermionic excitations as well as magnetic order. Thus this provides a possible way to understand the unconventional diffuse neutron scattering in rare-earth pyrochlores such as Yb$_2$Ti$_2$O$_7$, Er$_2$Sn$_2$O$_7$ and Er$_2$Pt$_2$O$_7$ at low/zero external magnetic fields. We calculate the dynamic spin structure factor to understand the nature of the diffuse two-particle continuum.
\end{abstract}
\date{\today}
\maketitle
\section{Introduction}  

Competition between magnetically ordered phases and quantum entangled  QSLs \cite{wen2017colloquium, PhysRevB.87.104406, anderson1973resonating,PhysRevLett.86.1881,wen2002quantum,kitaev2003fault,kitaev2006anyons,lee2008physics,balents2010spin, savary2016quantum} may be at the heart of understanding the low temperature magnetic properties of several rare-earth pyrochlore magnets, R$_2$T$_2$O$_7$.\cite{chern2018magnetic,PhysRevX.6.011034, PhysRevB.95.094422, PhysRevLett.82.1012,
PhysRevB.68.180401,PhysRevB.70.180404,applegate2012vindication,fennell2012power,lhotel2014first,pan2015measure,tokiwa2016possible,rau2018frustrated} In more than one of such systems, recent experiments reveal signatures of spin fluctuations characteristic to both magnetic order and possible fractionalized excitations expected in a QSL.  These systems then provide a natural context to investigate the interplay of the physics of spontaneous symmetry breaking and long range quantum entanglement in correlated condensed matter. Similar competition has been recently proposed to underlie the unconventional magnetic properties of the honeycomb lattice  QSL candidate-- RuCl$_3$.\cite{banerjee2016proximate}

The above interplay is embodied in  a series of neutron scattering experiments on  Yb$_2$Ti$_2$O$_7$\cite{PhysRevLett.119.057203, ross2011quantum, PhysRevB.93.064406,PhysRevB.92.064425,PhysRevLett.103.227202} which has attracted a lot of recent attention. In Yb$_2$Ti$_2$O$_7$, inelastic neutron scattering experiments  reveal sharp gapped magnon excitations at high magnetic fields ($\sim 5 $ T, such that the spins are polarized) which give way, at low fields ($\lesssim 0.5$ T), to low energy diffusive spin scattering around the Brillouin zone centre at an energy of $\sim0.15$ meV and extending all the way up to $1.5$ meV.\cite{PhysRevB.92.064425,PhysRevLett.119.057203} Similar effects were also reported in the THz spectroscopy experiments.\cite{pan2014low} This is in conjunction with other experiments which indicate a splayed ferromagnetic ground state for Yb$_2$Ti$_2$O$_7$ with transition temperature $\sim 0.2$ K\cite{doi:10.1143/JPSJ.72.3014,chang2012higgs,PhysRevB.93.064406,PhysRevB.93.144412, 0953-8984-28-42-426002,PhysRevLett.88.077204,PhysRevB.89.184416} that is drastically suppressed by application of magnetic field.\cite{PhysRevLett.119.057203,PhysRevB.93.144412} Equally interesting is Er$_2$Sn$_2$O$_7$ where magnetic order is observed below $\sim108$ mK.\cite{PhysRevB.97.024415,PhysRevLett.119.187202} However, even below the magnetic ordering temperature, a {\it quasi-elastic} spectral weight has been observed in the neutron scattering on powder samples.\cite{PhysRevLett.119.187202} Similarly, recent neutron scattering studies on powdered Er$_2$Pt$_2$O$_7$ suggest   magnetic order below $0.38$ K with unusual excitation spectra which cannot be accounted within linear spin-wave theory.\cite{PhysRevLett.119.187201} In a related material, Er$_2$Ti$_2$O$_7$, a delicate interplay of quantum fluctuation is a likely reason for the interesting {\it order-by-disorder} effects\cite{PhysRevLett.109.167201,PhysRevLett.109.077204} leading to a magnetically ordered state below $T=1.1$ K.\cite{PhysRevLett.101.147205} It has been suggested that the suppression of magnetic order in Er$_2$Sn$_2$O$_7$ and Er$_2$Pt$_2$O$_7$, in comparison to Er$_2$Ti$_2$O$_7$, is due to proximity to the phase boundary of competing magnetic orders.\cite{hallas2017experimental} Similarly, non-Kramers analogs such as Tb$_2$Ti$_2$O$_7$ and Pr$_2$Zr$_2$O$_7$ also show low temperature spin fluctuation dominated physics that poses interesting question regarding the nature of competing orders.\cite{0034-4885-77-5-056501} 
 
Several interesting ideas have been put forward to understand the above rich and diverse set of properties.  Analysis of the Hamiltonians appropriate for these systems in the classical limit\cite{PhysRevB.95.094422} reveal a rich phase diagram which places the systems such as Yb$_2$Ti$_2$O$_7$ and Er$_2$Sn$_2$O$_7$ near the boundary of two classical magnetically ordered phases. From this point of view,  the low energy anomalous  spin scattering, as found in the neutron experiments, originate from the competing order-parameter fluctuations near their mutual phase boundary.\cite{PhysRevB.95.094422,PhysRevB.92.064425} 

An alternate viewpoint starts by positing that the quantum fluctuations can stabilise a QSL in a regime where candidate magnetically ordered phases are energetically fragile due to competing interactions and the experimental observations for these materials should be understood in the lights of the competition between magnetically ordered phases and QSLs.  The central question then pertains to the relevant candidate QSLs for these materials. This idea has been  explored\cite{ross2011quantum,savary2012coulombic,lee2012generic} starting with a U(1) QSL with gapped bosonic electric and magnetic charges and gapless emergent photon  (the so called {\it quantum spin ice}\cite{PhysRevB.69.064404})  and more recently starting from a U(1) QSL (the so called {\it monopole flux state}\cite{PhysRevB.79.144432}) or a uniform Z$_2$ QSL, both with fractionalised fermionic quasi-particles. The U(1) monopole flux state also has a gapless photon whereas the uniform Z$_2$ has a gapped flux loop excitation.\cite{chern2018magnetic} These fractionalised quasiparticles naturally give rise to diffusive spin-structure factor through two-(quasi) particle continuum that contributes to the spin-spin correlations\cite{PhysRevLett.100.227201,PhysRevB.88.224413,punk2014topological,PhysRevB.86.224417,PhysRevLett.112.207203,knolle2018dynamics} with possible information of the statistics of the low energy quasi-particles.\cite{PhysRevLett.118.227201}

In this paper, we explore the feasibility of  stabilizing the time reversal invariant uniform $U(1)$ QSL in Hamiltonians applicable to rare-earth pyrochlores and examine their competition with possible magnetically ordered phases. Indeed such a QSL was proposed in context of rare-earth pyrochlores in Ref. \onlinecite{PhysRevX.6.011034}. Here, we explore the relevance of this uniform U(1) QSL with fermionic partons to the materials with particular emphasis on their instability to magnetically ordered phases. This, then provides an alternate starting point to understand the unconventional low energy magnetic fluctuations in several of these magnets.

 Due to the underlying long range quantum entanglement, the QSLs, in presence of symmetries, allows symmetry fractionalisation and long range interactions among emergent fermionic/bosonic (in three spatial dimensions) quasiparticles with fractionalised quantum numbers mediated by emergent $SU(2)$, $U(1)$ or Z$_2$ gauge fields.\cite{PhysRevB.87.104406,wen2002quantum}  Thus the low energy field theory of QSLs naturally takes the form of gauge theories coupled to gapped/gapless dynamic fermionic/bosonic matter fields and the QSLs represent the deconfined phase of such gauge theories. For the uniform $U(1)$ QSL under consideration, the low energy theory is described by fermionic quasiparticles (partons) coupled to a dynamic emergent $U(1)$ gauge theory. Starting from such a QSL with fermionic quasiparticles, {\it conventional} magnetic order can be obtained generically in a two-step  process -- (1) confinement of the fermionic partons to spin carrying gauge neutral spins, and (2) condensation of the spins. This immediately suggests the possibility of existence of an {\it unconventional} magnetically ordered phase where the partons are not confined.\cite{senthil1999quantum}

 A well known approach to capture basic features of a class of QSL including  the nature of the low energy quasi-particles, their symmetry transformation, as well as the nature of the emergent low energy gauge fluctuations are the parton mean-field theories\cite{wen2002quantum,baskaran1987resonating,wen2004quantum,PhysRevB.37.580,PhysRevB.37.3774,PhysRevB.38.5142,RevModPhys.78.17,PhysRevLett.62.1694,PhysRevB.42.4568,PhysRevLett.66.1773} where the spins are written in terms of bosonic or fermionic bilinears (see Eqs. \ref{eq_sploc} and \ref{eq_spglo}). These bosonic/fermionic  partons then transform under the projective representation of the symmetries leading to the projective symmetry group (PSG) based classification of QSLs realisable within the parton mean-field theories. \cite{wen2002quantum,wen2004quantum,PhysRevB.87.104406} The competition between the magnetically ordered phases and the QSLs can then be examined starting with the Curie-Weiss mean-field theories for the magnetic order and augmenting them with parton mean-fields. Self-consistent treatment of such generalised mean-field theories can then not only capture the magnetic phases and the QSLs, albeit at the mean-field level, but also allows for phases with co-existing magnetic and fractionalised excitations of the quantum ordered QSL. 

Our starting point is similar to that of Ref. \onlinecite{chern2018magnetic} which however starts from a different QSL-- the monopole flux state or an uniform $Z_2$ QSL and is also different from the gauge mean-field theory approaches with bosonic partons for Ref. \onlinecite{savary2012coulombic}. Our results from the self consistent mean-field calculations with a fermionic $U(1)$ QSL  ansatz with time reversal symmetry shows phases with both gapped and gapless parton band structures at different parts of the experimentally relevant phase diagram (Fig. \ref{fig_tbi}).   In the gapped region, the parton band structure is characterized by a non-zero $Z_2$ topological invariant and thus it realises a {\it fractionalised topological insulator}\cite{PhysRevB.85.224428} of the partons or the so called $(E_{fT}M_f)_\theta$ phase of Ref. \onlinecite{PhysRevX.6.011034}. Thus such a phase can support robust gapless surface states in addition to the gapped bulk fermion and gapless bulk photons as well as a gapped Dyon. We generically find that the QSL is unstable to small external magnetic fields. 

 Our mean field studies of the competition between the QSL and the ${\bf q}=0$ magnetic ordered phases reveal a rich phase diagram (Figs. \ref{fig:pd_dot} and \ref{fig_pd}) which supports fractionalized magnetically ordered phases (see Fig \ref{fig_cut}) in addition to the pure QSL phases and conventional magnetically ordered phases.  The fractionalized magnetically ordered phases are interesting from the point of view of the experiments on rare-earth magnets. The calculated dynamic structure factors show broad and diffused continuum (see fig.~\ref{fig_sf_pure_qsl} for example) resembling the low magnetic field neutron scattering in the experiments which gives way to sharper structures at higher magnetic fields once the QSL is destroyed. Our results then indicate an alternate starting point to understand the properties of rare-earth pyrochlore magnets such as Yb$_2$Ti$_2$O$_7$ and Er$_2$Sn$_2$O$_7$.

 The rest of the paper is organised as follows. We start in Sec. \ref{sec_rto} with a description of the physics of rare-earth pyrochlores and introduce the minimal spin Hamiltonian consistent with the symmetries of various phases of rare-earth pyrochlore magnets. For completeness, we briefly discuss the existing literature on the possible candidate QSLs as well as classical  magnetically ordered phases. In section \ref{sec:QSL} we introduce the uniform $U(1)$ QSL and discuss the parton mean-field description of such a state by introducing the appropriate bilinear channels for partons. We provide the self-consistent mean field calculations in an experimentally relevant regime for the materials Yb$_2$Ti$_2$O$_7$, Er$_2$Ti$_2$O$_7$, Er$_2$Sn$_2$O$_7$ and Er$_2$Pt$_2$O$_7$. The corresponding phase diagram is plotted in Fig. \ref{fig_tbi}. Within the mean-field theory we calculate the dynamic spin structure factor measured in inelastic neutron scattering experiments.  In Sec. \ref{section:magnetic_order} we focus on the magnetic instability of the uniform $U(1)$ QSL to the ${\bf q}=0$ magnetic orders that have been obtained in the classical limit. For which the corresponding phase diagram is plotted in fig. \ref{fig:pd_dot} and \ref{fig_pd}. We discuss the relevance of our calculations to the candidate materials such as Yb$_2$Ti$_2$O$_7$ and Er$_2$Sn$_2$O$_7$ in Sec. \ref{sec:Materials}.  We summarise our conclusions in Sec. \ref{sec:conclusion}. The details of different calculations are given in various appendices.

\section{The spin Hamiltonians for rare-earth pyrochlores}
\label{sec_rto}
\begin{figure}
\centering
\includegraphics[scale=0.3]{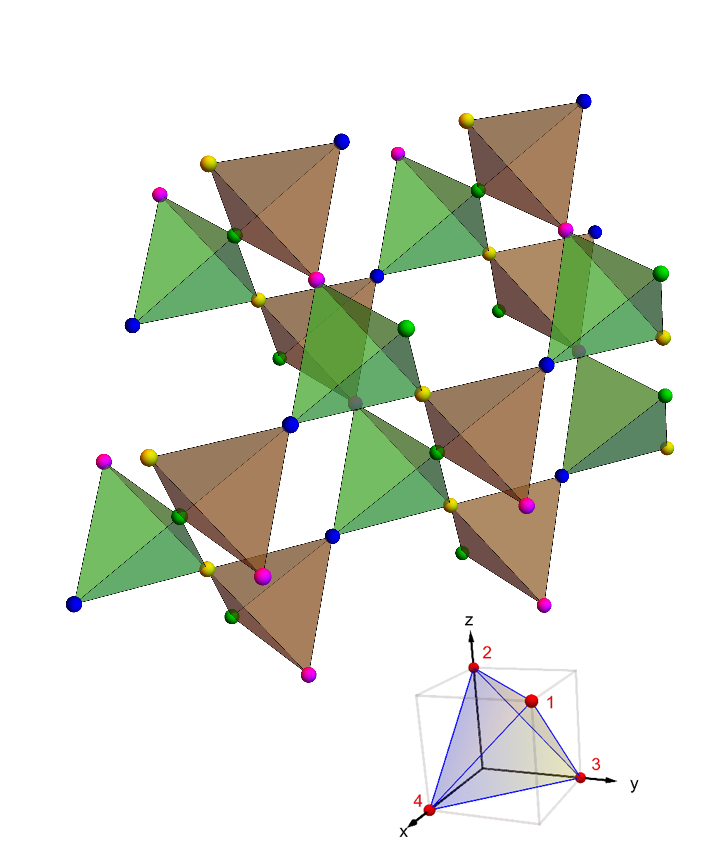}
\caption{(Top) The pyrochlore lattice formed by the rare-earth ions that carry effective spin-$1/2$ moments at each  vertices.  (Bottom) The unit cell containing four sites.}
\label{fig_pyro}
\end{figure}


Several rare-earth pyrochlore magnets with chemical composition R$_2$T$_2$O$_7$ (R = rare earth, T = transition metal) are now known.\cite{0034-4885-77-5-056501, gardner2010magnetic}  Magnetism arises due to the spin-orbit coupled ${\bf J=L+S}$ magnetic moments on the  rare-earth, R$^{3+}$, sites that form a network of corner-sharing tetrahedra, {\it i.e.}  the pyrochlore lattice, as shown in fig. \ref{fig_pyro}. These ${\bf J}$ moments are further split by local crystal field giving rise to effective doublets, {\it i.e.}, spin-$1/2$ pseudospins  which then interact among themselves in a highly anisotropic manner consistent with the symmetries of the pyrochlore lattice \cite{harris1997geometrical,gardner2010magnetic, PhysRevB.75.212404,onoda2011effective,PhysRevLett.119.057203} (see Appendix \ref{appen_ham_lattice} for details). For the rest of the work we shall concern ourselves with the pseudospins being Krammers doublets (as is the case for Yb$_2$Ti$_2$O$_7$ and Er$_2$Sn$_2$O$_7$)  satisfying the usual algebra  
$$ [s^\mu_i,s^\nu_j]=i\epsilon^{\mu\nu\lambda}s^\lambda_i$$  
and odd under time-reversal symmetry. The minimal Hamiltonian for the nearest neighbour spin exchange has the general form\cite{PhysRevB.75.212404, PhysRevB.95.094422,ross2011quantum}
\beq\label{eqn:globalH}
H_{global}=\sum_{\langle ij\rangle} \mathcal{J}^{\mu\nu}_{ij}s^\mu_is^\nu_j
\eeq

where $\langle ij\rangle$ indicates summation over nearest neighbour pairs of spin-1/2s (from hereon throughout the rest of the paper, we shall refer pseudospins as spins),  $s_i^\mu$ $(\mu=x,y,z)$, on the sites of the pyrochlore lattice.  $\mathcal{J}^{\mu\nu}_{ij}(=\mathcal{J}^{\nu\mu}_{ji})$ represents a $3\times 3$ matrix for a given bond $\langle ij\rangle$ that represents the coupling strengths\cite{PhysRevB.75.212404, PhysRevB.95.094422,ross2011quantum} the form of which is given in Appendix \ref{appen_ham_lattice}. On a given bond the coupling constant matrices have four independent symmetry allowed parameters\cite{PhysRevB.78.094418} denoted by $J_1, J_2, J_3$ and $J_4$ (see Appendix \ref{Jglobal}). Given the structure of $\mathcal{J}^{\mu\nu}_{ij}$ on any one bond, it is completely determined by symmetries on all the other bonds.

Due to the interplay of the strong spin-orbit coupling and the fact that the local symmetry axis for the D$3d$ crystal field is different at the different sublattices, the spins have a natural axis of quantization - the local $[111]$ direction of the site of the pyrochlore lattice. With the spins defined along the local quantization direction (see Appendix \ref{appen_ham_lattice}), the Hamiltonian in eqn. \ref{eqn:globalH} can be re-written as,\cite{ross2011quantum,onoda2011effective}
\bea\mylabel{eqn:localH}
H_{local}&=& \sum_{\langle ij \rangle} \big( J_{zz} S_i^z S_j^z - J_{\pm}(S_i^+ S_j^- + S_i^- S_j^+)\nonumber \\
&+& J_{\pm \pm} (\gamma_{ij} S_i^+ S_j^+ +\gamma_{ij}^* S_i^- S_j^-)\nonumber \\
&+& J_{z\pm} \big[S_i^z(\zeta_{ij} S_j^+ +\zeta_{ij}^* S_j^-) + i \leftrightarrow j\big] \big)
\eea
with $\zeta_{ij}$ being given by eqn. \ref{zeta} with $\gamma_{ij}=-\zeta_{ij}^*$ and $S^\alpha_i$ denotes the spin components defined in the local quantisation basis. The relation between the coupling constants in the global and the local descriptions\cite{ross2011quantum} are given in eqn. \ref{eq_glob_loc}.  We shall alternatively use the local and global basis to make connections with existing literature as well as explain our results.

An experimentally relevant classification of rare-earth pyrochlore magnets is based on the relevance of the terms in the Hamiltonian in the local axis given by Eqn. \ref{eqn:localH}.  For materials such as R$=$ Ho, Dy,\cite{0034-4885-77-5-056501,Fennell16102009,PhysRevB.79.014408} the non-commuting ``quantum" terms other than $J_{zz}$ are negligible giving rise to the fascinating physics of classical spin-ice (however in such compounds long range dipolar interactions are prominently present) with emergent magnetic monopoles.\cite{PhysRevB.57.R5587,PhysRevLett.93.167204,castelnovo2008magnetic, castelnovo2012spin, bramwell2009measurement,Morris411,PhysRevB.71.014424,henley2010coulomb} Perturbations about this classical Ising limit by terms such as   $J_\pm$ lead to the physics of {\it quantum spin-ice} which is a three dimensional U(1) QSL whose low spectrum consists, in addition to a gapped bosonic monopole and gapless photon, a gapped bosonic electric charge.\cite{PhysRevB.69.064404} All these emergent excitations, particularly the low energy emergent photon, are highly non-local collective modes in terms of the underlying spins. A combination of field theoretic and numerical analysis presently yields quite convincing evidence for the existence of such an U(1) QSL in the vicinity of the classical spin-ice\cite{PhysRevB.69.064404,PhysRevLett.100.047208,PhysRevLett.91.167004,PhysRevB.68.184512,lee2012generic,savary2012coulombic,PhysRevLett.108.067204,PhysRevB.90.214430,PhysRevB.86.075154}, in the regime where $J_{zz}$ is the biggest energy scale.

However, it is now known from various experimental estimates of coupling constants that for several of rare-earth magnets including Yb$_2$Ti$_2$O$_7$, Tb$_2$Ti$_2$O$_7$, Er$_2$Ti$_2$O$_7$ and Er$_2$Sn$_2$O$_7$  \cite{PhysRevLett.106.187202,0953-8984-21-49-492202,1742-6596-320-1-012065} (see Appendix \ref{appen_coupling}), $J_{zz}$ is not the largest energy scale and gauge mean-field theory approaches have been applied\cite{ross2011quantum,savary2012coulombic,lee2012generic}  to study the the quantum spin ice as well as its competition with magnetically ordered phases in an extended parameter regime. These calculations reveal a rich phase diagram that includes, apart from extended regimes of quantum spin ice and conventional magnetically ordered phases, a {\it Coulomb ferromagnet} which can support both magnetic order as well as fractionalised excitations. In a very different, but interesting, parameter regime, the nearest neighbour spin-$1/2$ isotropic Heisenberg antiferromagnet on the pyrochlore lattice is also expected to have a QSL ground state due to the high level of frustration.\cite{PhysRevLett.80.2933,PhysRevB.79.144432} While presently we do not know for certain the exact nature of the QSL, fermionic parton mean field theory and variational Monte-Carlo calculations favour a U(1) QSL with {\it monopole flux}.\cite{PhysRevB.79.144432} However, the rare-earth pyrochlores are described by anisotropic spin Hamiltonians and it is not clear that the monopole flux state is also the natural QSL candidate (see Ref. \onlinecite{chern2018magnetic} for results starting with the monopole flux state for rare earth pyrochlores). 

In parallel, efforts to understand the classical phase diagram for the above Hamiltonian have revealed an intricate competition among different magnetically ordered phases. These studies have mostly used the global basis (eqn. \ref{eqn:globalH}) and in particular the $J_3=-1$ and $J_4=0$ hyperplane, relevant to Yb$_2$Ti$_2$O$_7$, Er$_2$Ti$_2$O$_7$ and Er$_2$Sn$_2$O$_7$  have been thoroughly investigated\cite{PhysRevB.95.094422} in search of magnetic orders with lattice translation symmetry-- {\it i.e.} ${\bf q}=0$ magnetic orders. These magnetic orders, thus break lattice point group symmetries ({\it i.e.}, symmetries of a single tetrahedron) along with time reversal.  As reviewed in section \ref{section:magnetic_order}, in this hyperplane the relevant ${\bf q}=0$ magnetically ordered phases are-- (1) a non-collinear ferromagnetic (splayed ferromagnet) phase with net magnetization along $[001]$ direction and spin canting away from this axis in a staggered manner, (2) antiferromagnetic phases transforming under a doublet $\mathrm{E}$, representation of tetrahedral group ($\mathcal{T}_d$) (non-coplanar atiferromagnetic phase $\Psi_2,$ and coplanar antiferromagnetic phase $ \Psi_3$), and (3) the Palmer-Chalker phase\cite{PhysRevB.62.488}, a coplanar antiferromagnetic phase with spins lying in [001] plane which transforms under a triplet ($\text{T}_2$) representation of $\mathcal{T}_d$. The dotted lines in the Fig. \ref{fig_pd} show the boundaries between the magnetic phases on the $J_1-J_2$ hyper-plane (with $J_3=-1$ and $J_4=0$).  These studies place  compounds such as Yb$_2$Ti$_2$O$_7$  near the phase boundary of the splay ferromagnet and the antiferromagnetic phase while Er$_2$Ti$_2$O$_7$ and Er$_2$Sn$_2$O$_7$ near the boundary of the antiferromagnetic and Palmer Chalker phase (see Fig. \ref{fig_pd}).

Given the current experimental observations, particularly the neutron scattering results, it is interesting to ask if the origin of the diffuse spin scattering stems from the proximity to a QSL and if so, what type of QSL can be realized in these materials ? The fact that in many of these compounds the transverse exchanges are sizable (in fact in some cases larger than $J_{zz}$) and yet they are far from the isotropic Heisenberg limit, opens a room for trying to understand the experimental results starting from a QSL different from quantum spin-ice and investigate for magnetic instability to ascertain  their relevance to capture the general phenomenology of this class of rare earth pyrochlore magnets. For example, recent neutron scattering experiments on Yb$_2$Ti$_2$O$_7$\cite{PhysRevLett.119.057203} shows a sizable spectral weight at very low energies near the $\Gamma$-point of the Brillouin zone at low external magnetic fields. Within the present understanding, natural QSL candidates that can account for such low energy spectral weight in the dynamic spin structure factor consists gapless fermionic quasi-particles.
There are many QSLs that can be classified partially within the projective symmetry group based classifications.\cite{PhysRevLett.119.057203,PhysRevB.87.104406} Here, however, we shall consider the simplest of such QSLs within the fermionic parton mean-field theory, the uniform U(1) QSL and understand its features as well as its instability to the magnetic ordered phases found in the classical limit within a parton mean field theory. We will restrict ourselves to the $J_3=-1$ and $J_4=0$ plane of the phase diagram. The results are summarised in Figs. \ref{fig_tbi} and \ref{fig_pd}. 
\section{Uniform U(1) QSL with fermionic partons on pyrochlore lattice}
\mylabel{sec:QSL}
The possibility of more than one $U(1)$ QSLs on a pyrochlore lattice with low energy fermionic partons have been investigated previously within parton mean field theories.\cite{chern2018magnetic,PhysRevB.79.144432,PhysRevB.85.224428,PhysRevB.89.075128,PhysRevB.78.180410}  The variational Monte-Carlo calculations on the spin-rotation symmetric nearest neighbour Heisenberg antiferromagnet find the so called {\it monopole} flux state to be of minimum energy among a handful of fermionic parton ansatz.\cite{PhysRevB.78.180410} This result has been extended to the above Hamiltonian (eqn. \ref{eqn:globalH}) very recently in Ref. \onlinecite{chern2018magnetic} (which also examined the uniform $Z_2$ QSL).
However, the present experimental phenomenology begs for further understanding of the candidate QSLs for this family of materials. Here, we shall start with the simplest $U(1)$ QSL\cite{PhysRevX.6.011034} with fermionic partons on the pyrochlore lattice as the alternate starting point to examine the competition between this QSL and various magnetic orders. To this end, we start by understanding the properties of the uniform U(1) QSL with fermionic partons. 

\subsection{The fermionic parton mean field theory}
\mylabel{sec:parton}

The parton mean-field theory provides a systematic starting point to develop  the low energy effective theories for QSL. To this end we define fermionic parton annihilation operators in the local and global basis by $\{f_{i\uparrow},f_{i\downarrow}\}$ and $\{F_{i\uparrow},F_{i\downarrow}\}$ respectively. 
 The fermionic partons in the global and local basis are related by $f_{i\sigma}=V_{i;\sigma\sigma'}F_{i\sigma'}$ (see Appendix \ref{appen_loc_glob}). The corresponding spin operators, bilinears of the fermions, are given by
\begin{align}
S_i^a=\frac{1}{2}f_{i\alpha}^\dagger  [\sigma^a]_{\alpha \beta} f_{i\beta}~~~~{\rm (in~the~local ~basis)}
\label{eq_sploc}
\end{align}
and
\begin{align}
s_i^a=\frac{1}{2}F_{i\alpha}^\dagger  [\sigma^a]_{\alpha \beta} F_{i\beta}~~~~{\rm (in~the~global ~basis)},
\label{eq_spglo}
\end{align}
where $\sigma^a~ (a=x,y,z)$ are Pauli matrices and $\alpha,\beta =\uparrow,\downarrow$. For a faithful representation of the Hilbert space the spinon operators satisfy the constraint 
\begin{align}
\sum_{\alpha} f_{i\alpha}^\dagger f_{i\alpha}=\sum_{\alpha} F_{i\alpha}^\dagger F_{i\alpha}=1
\label{eq_constraint}
\end{align}
per site.

 The bilinear spin-spin interaction terms in the spin Hamiltonian (\eqn{eqn:localH}) becomes quartic in terms of the parton operators.  In order to perform a parton mean-field analysis these quartic parton terms are decoupled into various channels which are quadratic in terms of the partons. Due to the lack of spin rotation symmetry, we need to consider both the singlet and triplet hopping (particle-hole) and pairing (particle-particle) channels, given by\cite{PhysRevB.80.064410,PhysRevB.86.224417,PhysRevB.85.224428} 

\bea\mylabel{eqn:MFTchannels}
\chi_{ij}&=&f_{i\alpha}^\dagger \delta_{\alpha \beta} f_{j\beta}~~~~~~~~~~~\widetilde{\chi}_{ij}=F_{i\alpha}^\dagger \delta_{\alpha \beta} F_{j\beta} \nonumber \\
E_{ij}^a&=& f_{i \alpha}^\dagger [\tau^a]_{\alpha \beta} f_{j\beta}~~~~~~~\widetilde{E}_{ij}^a= F_{i \alpha}^\dagger [\tau^a]_{\alpha \beta}F_{j\beta}\nonumber \\
\eta_{ij}^a&=& f_{i \alpha}[i \tau^2]_{\alpha \beta} f_{j\beta}~~~~~~~\widetilde{\eta}_{ij}^a= F_{i \alpha}[i \tau^2]_{\alpha \beta} F_{j\beta}\nonumber \\
D_{ij}^a&=& f_{i \alpha} [i\tau^2 \tau^a]_{\alpha \beta} f_{j\beta}~~~~\widetilde{D}_{ij}^a= F_{i \alpha} [i\tau^2 \tau^a]_{\alpha \beta} F_{j\beta}.
\eea

The above parton parametrization of the spins is gauge redundant with a $SU(2)$ gauge group\cite{PhysRevB.80.064410} which is generally broken down to $Z_2$ when all the above mean-field decoupling channels are present.\cite{wen2002quantum,baskaran1987resonating,wen2004quantum,PhysRevB.37.580,PhysRevB.37.3774,PhysRevB.38.5142,RevModPhys.78.17,PhysRevLett.62.1694,PhysRevB.42.4568,PhysRevLett.66.1773} However we are interested in $U(1)$ mean field ansatz where all the particle-particle channels are set to zero whence the resultant gauge-group is $U(1)$ where the gauge transformation on the local basis (say) are defined by 
\begin{align}
f_{i\alpha}\rightarrow e^{i\theta_i}f_{i\alpha},
\label{eq_gauge}
\end{align}
 where $\theta_i\in (0,2\pi]$. Clearly the spin operators in eqn. \ref{eq_sploc} are gauge invariant. Thus the partons carry electric charge of the emergent U(1) gauge field that gains dynamics on interacting out the high energy parton modes. However, in our mean-field treatment we shall ignore the dynamics of the gauge field and their coupling to low energy partons.  

Keeping only the particle-hole channels, the Hamiltonian in the local basis can then be derived from \eqn{eqn:localH}, in terms of the singlet and triplet hopping terms ($\chi_{ij},E_{ij}^x,E_{ij}^y,E_{ij}^z$) and is given by 
\beq\mylabel{eqn:Echi_hamiltonian_1}
H_{local}=\sum_{\langle ij\rangle} H_f^{ij},
\eeq 
where
\bea\mylabel{eqn:Echi_hamiltonian_2}
H_f^{ij}&=& \frac{2J_{zz}}{16} \big[ -\chi_{ij}^\dagger \chi_{ij}-E_{ij}^{z\dagger} E_{ij}^z +E_{ij}^{x\dagger} E_{ij}^x+E_{ij}^{y\dagger} E_{ij}^y\big]\nonumber \\
&&- \frac{8 J_{\pm}}{16}\big[ -\chi_{ij}^\dagger \chi_{ij}+E_{ij}^{z\dagger} E_{ij}^z\big]\nonumber \\
&&+ \frac{4 J_{\pm \pm}}{16} \left[-(\gamma_{ij} +\gamma_{ij}^*)E_{ij}^{x\dagger} E_{ij}^{x} +(\gamma_{ij} +\gamma_{ij}^*)E_{ij}^{y\dagger} E_{ij}^{y}\right.\nonumber\\
&&\left. -i(\gamma_{ij} -\gamma_{ij}^*)E_{ij}^{y\dagger} E_{ij}^{x} -i(\gamma_{ij} -\gamma_{ij}^*)E_{ij}^{x\dagger} E_{ij}^{y} \right]\nonumber \\
&&-\frac{4J_{z\pm}}{16} \left[(\zeta_{ij} +\zeta_{ij}^*)E_{ij}^{z\dagger} E_{ij}^{x} +(\zeta_{ij} +\zeta_{ij}^*)E_{ij}^{x\dagger} E_{ij}^{z} \right.\nonumber\\
&&\left.+i(\zeta_{ij} -\zeta_{ij}^*)E_{ij}^{z\dagger} E_{ij}^{y} +i(\zeta_{ij} -\zeta_{ij}^*)E_{ij}^{y\dagger} E_{ij}^{z}  \right].\nonumber\\
\eea
Similar expressions can be derived for the decoupling in the global basis starting with eqn. \ref{eqn:globalH}. From this, the mean-field Hamiltonian is obtained by keeping the following mean-field variables.
\bea
\bar\chi_{ij}&=&\langle f_{i\alpha}^\dagger \delta_{\alpha \beta} f_{j\beta}\rangle~~~~~~~~~~~\bar{\widetilde{\chi}}_{ij}=\langle F_{i\alpha}^\dagger \delta_{\alpha \beta} F_{j\beta}\rangle \nonumber \\
\bar E_{ij}^a&=& \langle f_{i \alpha}^\dagger [\tau^a]_{\alpha \beta} f_{j\beta}\rangle~~~~~~~\bar{\widetilde{E}}_{ij}^a= \langle F_{i \alpha}^\dagger [\tau^a]_{\alpha \beta}F_{j\beta}\rangle.
\label{eq_qsl_mf}
\eea

\subsection{Symmetry considerations}
\label{subsection:symmetry}

Under the above gauge transformation (eqn. \ref{eq_gauge}), the mean field variables  transform as
\begin{align}
\bar\chi_{ij}\rightarrow\bar\chi_{ij}e^{i(\theta_i-\theta_j)},~~~~\bar E^a_{ij}\rightarrow \bar E^a_{ij}e^{i(\theta_i-\theta_j)}.
\end{align}
Thus they are not gauge invariant. Indeed, on integrating out high energy partons, the gauge field becomes dynamical and gives rise to two (for different polarisation) gapless photon modes. These are nothing but phase fluctuations of $\bar\chi_{ij}$ and $\bar E^a_{ij}$.  However, since the partons carry electric charge of the gauge field, they transform under projective representation under various symmetries of the system--{\it i.e.} lattice symmetries and time reversal\cite{wen2002quantum,wen2004quantum,PhysRevB.87.104406} such that all symmetry transformations must be augmented by possible gauge transformation. Indeed different projective representation of the symmetries can lead to different types of QSLs. The uniform $U(1)$ QSL is the simplest of them where all the gauge transformations are trivial, or in other words all the symmetries are present manifestly.  This is then a different QSL from both the monopole flux $U(1)$ QSL and the uniform $Z_2$ QSL and provides an alternate starting point to understand the physics of the rare earth pyrochlore magnets. This highly constrains the structure of the parton mean field theory as we discuss below.

Pyrochlore lattice is a network of corner sharing tetrahedra each of which form a four site unit cell arranged on an underlying face centered cubic Bravais lattice (see fig. \ref{fig_pyro} and appendix \ref{appen_ham_lattice}).  The point group is an octahedral group, $O_h\equiv\mathcal{T}_d \times \mathcal{I}$. Here $\mathcal{T}_d $ is the tetrahedral group and $\mathcal{I}$ is the group of inversion.\cite{PhysRevB.96.125145} (See Appendix \ref{appen_symmetries} for details).  In addition to the lattice symmetries, the system also has time reversal invariance in absence of the external magnetic field. As mentioned before, the spins, being Kramers doublet are odd under time reversal.

For manifestly time reversal invariance, we must have 
\begin{align}
\bar{\chi}\in \Re~~~{\rm and}~~~ \bar{E}_z,\bar{E}_x, \bar{E}_y\in \Im.
\end{align}
Further, lattice symmetry relations imply (see Appendix \ref{appen_symmetries}) that there are only two independent parameters, namely $E^x$ and $E^z$, in local coordinates.  This is exactly same as the statement that in presence of SOC, the tight-binding model for electrons on the pyrochlore lattice contains two independent hopping parameters.\cite{PhysRevB.85.045124,PhysRevB.87.214416}

\subsection{The Mean field Hamiltonian}
\label{pure_qsl_minimization}

\begin{table}
\begin{tabular}{ | c | c | c | c |  c |}
\hline
     $J_1$~ & 
     $J_2$~ & 
    $\bar{E}_z$~ & 
    $\bar{E}_x$~ &
    $E_{MF}$ \\
\hline
\hline
$3.0$ & $-2.0$ & $-0.41$ & $0.00$  & $-0.247$ \\
\hline
$0.0$ & $-2.0$ & $-0.25$ & $0.21$ & $-0.265$\\
\hline
$1.0$ & $1.0$ & $-0.33$ & $-0.08$ & $-0.128$\\
\hline
$2.0$ & $1.75$ & $-0.18$ & $-0.14$ & $-0.240$\\
\hline
$-2.0$ & $-2.0$ & $-0.31$ & $0.18$  & $-0.291$ \\
\hline
$-1.0$ & $-3.0$ & $-0.29$ & $0.19$ & $-0.356$\\
\hline
$3.0$ & $3.0$ & $-0.07$ & $-0.18$ & $-0.388$\\
\hline
\end{tabular}
\caption{Independent mean field parameter values in the local basis($\bar{E}_z$ and $\bar{E}_x$, in Col. 3 and 4 respectively) obtained by minimizing the mean-field energy per site as in \eqn{eqn:EMF}. The mean-field energy minima values per site ($E_{MF}$) are given in Col. 5. Results are shown for the parameters values $J_1$ and $J_2$ as given in Col.1 and 2 respectively, keeping the parameters $J_3=-1.0, J_4=0.0$ fixed. The first four rows in this table are for the points $a,b,c,d$ in  fig. \ref{fig_tbi}.}
\label{QSL_params}
\end{table}

The mean field Hamiltonian, expressed in terms of the independent mean field parameters in the local basis is given in \eqn{eqn:HMF}, where $E^z \equiv E_{12}^z,~E^x \equiv E_{12}^x$

\begin{align}\label{eqn:HMF}
H_{MF}=&\frac{3N}{16}\big[2(J_{zz}+4J_{\pm})|\bar{E}^z|^2 +8(-J_{zz}+4J_{\pm\pm})|\bar{E}^x|^2 \big]\nonumber\\
+&\frac{1}{16} 3 N \big[ 16 J_{z\pm}( \bar{E}^{z\dagger} \bar{E}^{x} + \bar{E}^{x\dagger} \bar{E}^{z})] \nonumber\\
+& \sum_{\bf{k}\in BZ} f_a^{\sigma_1\dagger}(\textbf{k}) H^{ab}_{\sigma_1 \sigma_2}(\textbf{k}) f_b^{\sigma_2}(\textbf{k}),
\end{align}
where $N$ is the total number of sites in the pyrochlore lattice, ${a,b} \in 1,2,3,4$ denote the sub-lattice indices as shown in fig. \ref{fig_pyro} and $\sigma_1,\sigma_2 \in \{ \uparrow ,\downarrow \}$ denote the spin indices. The forms of $H^{ab}_{\sigma_1 \sigma_2}$ are given in Appendix \ref{appen_MFT_details}. 

The mean-field ground-state is obtained by filling in the single particle energies for a parton band-structure where the mean-field parameters $\bar{E}^x$ and $\bar{E}^z$ (in local basis, say) are determined self-consistently with the constraint of one parton per site (\eqn{eq_constraint}) implemented on average using a chemical potential, $\mu$, which is also determined self-consistently. This is done by minimising the mean-field energy per site $E_{MF}$ as given in \eqn{eqn:EMF} in the local basis, the $\mu$ is determined at each step of the minimization by solving \eqn{eqn:mu}.
\bea\mylabel{eqn:EMF}
E_{MF}&=& \frac{3}{16} \big[ 2(J_{zz}+4J_{\pm})|\bar{E}^z|^2 +8(-J_{zz} +4J_{\pm\pm})|\bar{E}^x|^2 \big]\nonumber\\
&&+\frac{3}{16}\big[ 16 J_{z\pm}( \bar{E}_{12}^{z\dagger} \bar{E}_{12}^{x} + \bar{E}_{12}^{x\dagger} \bar{E}_{12}^{z})]\\ \nonumber
&+& \frac{1}{4 N_{UC}} \sum_{{\bf k} \in BZ} \sum_{i=1}^{N_{band}}  \Lambda_i(\textbf{k}) \Theta\big[\mu,\Lambda_i(\textbf{k}) \big].
\eea

In \eqn{eqn:EMF} and \eqn{eqn:mu}, $\Lambda_i$~s are the eigenvalues of $H^{ab}_{\sigma_1 \sigma_2}$, $N_{band}(=8)$ is the number of energy-bands for the pyrochlore lattice. $N_{UC}(=N/4)$ is the number of unit cells in the system and $\Theta(a,x)$ is the Heaviside theta function i.e. if $x<a, \Theta(a,x)=1$, else $\Theta(a,x)=0$.  The chemical potential is determined by the condition
\beq\mylabel{eqn:mu}
\frac{2}{N_{band}N_{UC}}\sum_{{\bf k} \in BZ} ~\sum_{i=1}^{N_{band}}~\Theta\big[\mu,\Lambda_i(\textbf{k}) \big]=1.
\eeq
Further details of the minimisation procedure is given in Appendix \ref{appen_mft_min}. The  values of the independent mean field parameters in the local basis and corresponding mean field energy per site is given in TABLE \ref{QSL_params} for some representative points in the phase diagram.

\subsection{The uniform U(1) QSL}
\mylabel{sec:MFT}

\begin{figure}
\centering
\includegraphics[scale=0.645]{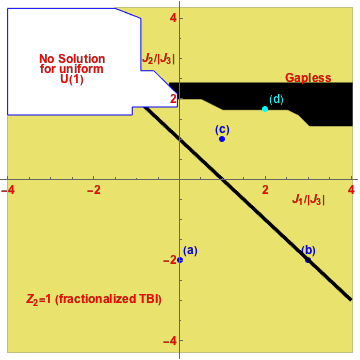}
\caption{Two uniform $U(1)$ QSLs are obtained by minimizing $E_{MF}$ in \eqn{eqn:EMF} with respect to the two independent parameters $E^z, E^x$(listed in Table \ref{QSL_params} for some sample points). The yellow (black) region corresponds to the QSL with gapped (gapless) parton excitations arising from complete (partial) filling of the parton bands. In the gapped phase the partons form a strong topological band insulator with $Z_2$ indices given by $[1;000]$. The points marked as $a,b,c,d$ represents points for which the patron band structure is plotted in fig. \ref{fig_bs_pure_qsl}. In the top left corner the white region marks region of no stable(conditions of stability is given in Appendix~\ref{appen_mft_min}) solution for uniform $U(1)$ QSL ansatz in the singlet and triplet hopping channels.}
\label{fig_tbi}
\end{figure}

\begin{figure}
\centering
\subfigure[$J_1=3.0$; $J_2=-2.0$]{
\includegraphics[width=0.45\linewidth,height=3.0cm]{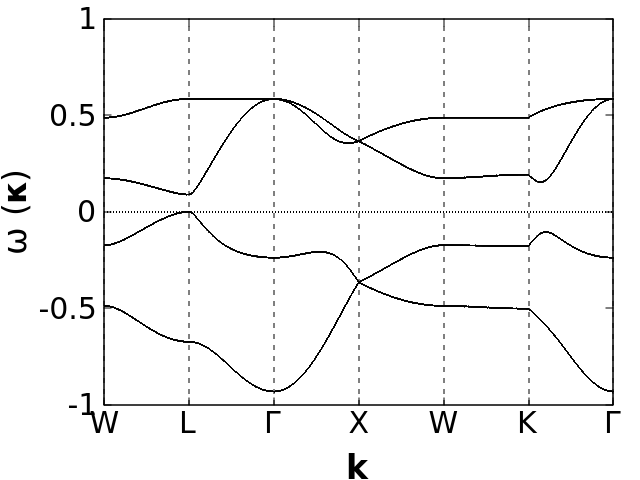}
}
\subfigure[$J_1=0.0$; $J_2=-2.0$]{
\includegraphics[width=0.45\linewidth,height=3.0cm]{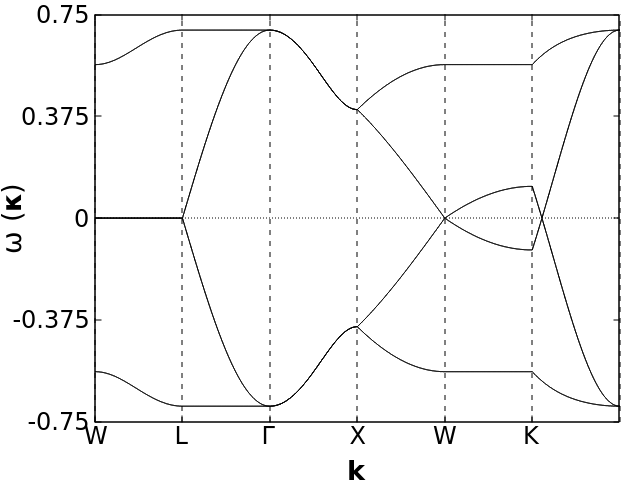}
}
\subfigure[$J_1=1.0$; $J_2=1.0$]{
\includegraphics[width=0.45\linewidth,height=3.0cm]{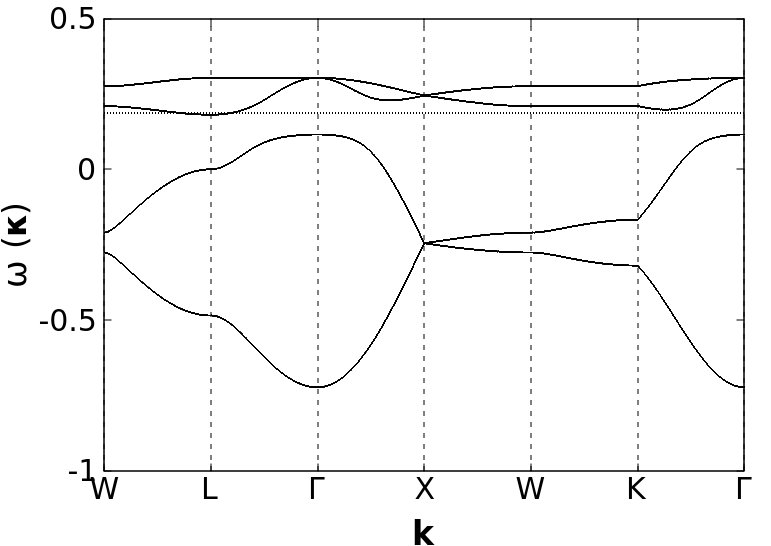}
}
\subfigure[$J_1=2.0$; $J_2=1.75$]{
\includegraphics[width=0.45\linewidth,height=3.0cm]{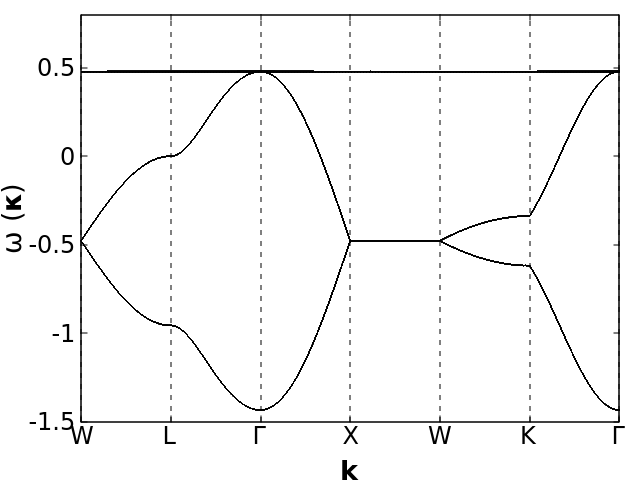}
}
\caption{Representative fermionic parton band structures for U(1) uniform QSL ansatz with triplet and singlet hopping channels, for the points $a,b,c,d$ of fig. \ref{fig_tbi}. The independent mean field parameters obtained by minimizing $E_{MF}$ in \eqn{eqn:EMF} are listed in Table.~\ref{QSL_params}. The dotted line in each figures denotes the chemical potential. Figs. \ref{fig_bs_pure_qsl}$(a)$, and \ref{fig_bs_pure_qsl}$(d)$ shows gapless excitations (in the black region of fig. \ref{fig_tbi}) and Figs. \ref{fig_bs_pure_qsl}$(b)$ and \ref{fig_bs_pure_qsl}$(c)$ shows gapped excitations (in the yellow regions of fig. \ref{fig_tbi}).}
\label{fig_bs_pure_qsl}
\end{figure}

The above calculations, depending on the parameter regime yields two types of uniform U(1) QSLs - gapped and gapless in most part of the $(J_3=-1, J_4=0)$ hyper-plane except for a part of the quadrant where $J_1<0$ and $J_2>0$. The phase diagram is given in fig. \ref{fig_tbi}. The representative parton band structures are plotted in fig. \ref{fig_bs_pure_qsl}. Due to the inversion symmetry being explicitly present, the bands are doubly degenerate with the right degeneracies at the different high symmetry points\cite{PhysRevB.87.214416} in both the gapped and the gapless regimes. At the mean field level, the lower half of the states are filled. The gapless band structures are characterised by small band-touching or more generally Fermi pockets that will have important consequence for the spin structure factor (see below). Due to the presence of the triplet decoupling channels, bond nematic (since the real space and spin space are coupled here it is a spin-orbital bond nematic) order is present. \cite{PhysRevB.80.064410,PhysRevB.85.224428} Indeed, we find that uniaxial nematic order parameter is non-zero. However the director of the nematic is perpendicular to every bond and hence does not break any symmetry of the system. Fluctuations of the director are gapped and are related to the amplitude fluctuations of the triplet decoupling channels. However the phase fluctuations of the triplet channels are related to two gapless emergent U(1) photon modes. 
\paragraph*{Fractionalised topological insulator\cite{PhysRevB.85.224428}:} For the gapped phase, the fermionic parton band structure is characterised by a non-zero $Z_2$ topological invariant. In fact, in this regime, the partons form a strong topological band insulator\cite{PhysRevB.76.045302} like its electron counterpart.\cite{PhysRevB.85.045124,PhysRevB.87.214416} This phase is exactly the fractionalised topological insulator found in Ref. \onlinecite{PhysRevB.85.224428} or the $(E_{fT}M_f)_\theta$ phase discussed in the same context in Ref. \onlinecite{PhysRevX.6.011034}. However, compared to Ref. \onlinecite{PhysRevB.85.224428}, in the present case, since there is no spontaneously broken spin rotation symmetry, there are no gapless Goldstone modes. However there are two gapless photon modes in the bulk in addition to the gapless surface states of the partons protected by the $Z_2$ invariant. Since the emergent gauge theory is compact, it allows a gapped monopole excitation. However the monopole gains $1/2$ electric charge due to the {\it Witten effect}\cite{witten1979dyons} arising from the topological band-structure of the partons.
\subsection{Dynamic spin Structure factor for the uniform U(1) QSL}

\begin{figure}[!htp]
\centering
\subfigure[$J_1=3.0$; $J_2=-2.0$]{
\includegraphics[width=0.45\linewidth,height=3.0cm]{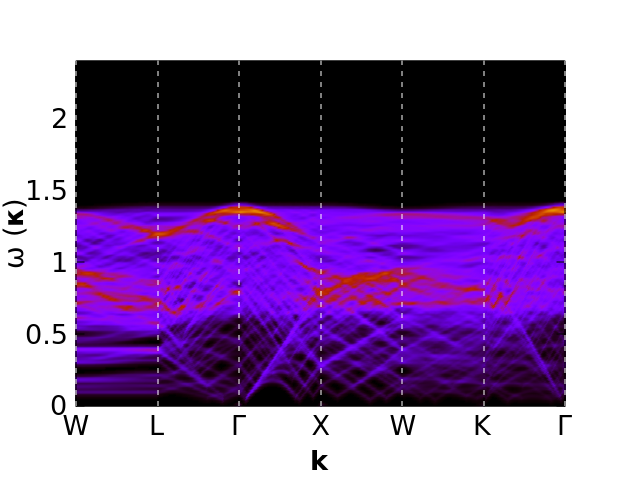}
}
\subfigure[$J_1=0.0$; $J_2=-2.0$]{
\includegraphics[width=0.45\linewidth,height=3.0cm]{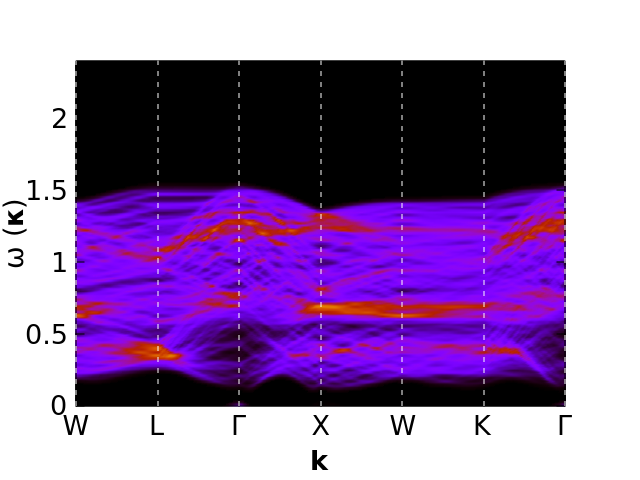}
}
\subfigure[$J_1=1.0$; $J_2=1.0$]{
\includegraphics[width=0.45\linewidth,height=3.0cm]{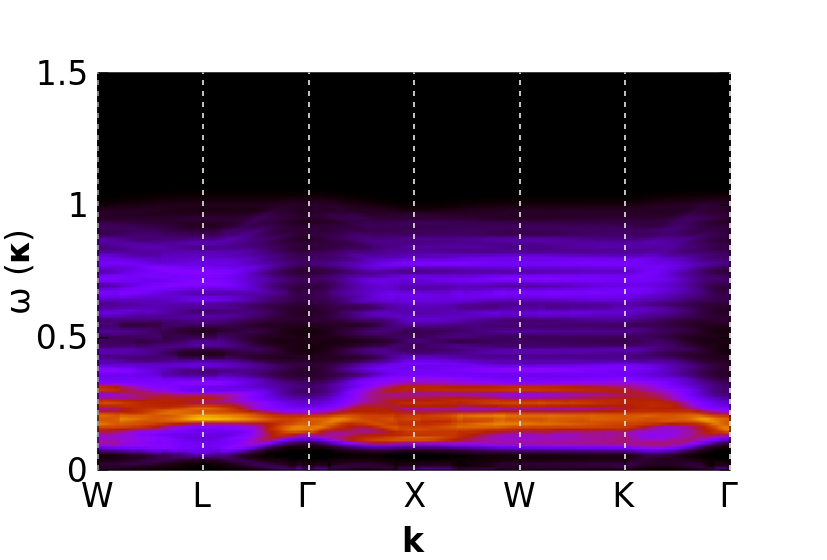}
}
\subfigure[$J_1=2.0$; $J_2=1.75$]{
\includegraphics[width=0.45\linewidth,height=3.0cm]{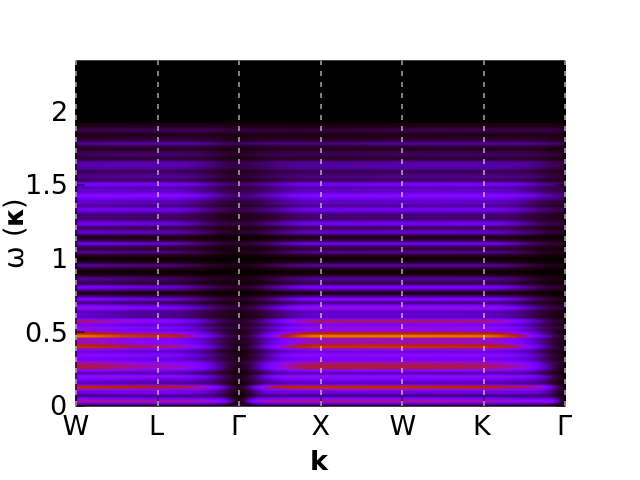}
}
\caption{Representative dynamical structure factors for U(1) uniform QSL ansatz with triplet and singlet hopping channels, for the points $a,b,c,d$ of fig. \ref{fig_tbi} and corresponding band structures in fig. \ref{fig_bs_pure_qsl}. }
\label{fig_sf_pure_qsl}
\end{figure}

Neutron scattering probes the spin correlations in the system through the dynamic spin structure factor as
\beq\mylabel{eqn:SF1}
\Omega(\textbf{Q},\omega)=\sum_{\alpha,\beta} (\delta_{\alpha \beta}-\widehat{k}_\alpha \widehat{k}_\beta)\chi^{\alpha \beta}(\textbf{Q},\omega).
\eeq
Where the dynamic spin structure factor is given by 
\beq
\chi^{\alpha \beta}(\textbf{Q},\omega)=\sum_{a,b}\langle M_a^{\alpha}(-\mathbf{Q},-\omega) M_b^{\beta}(\mathbf{Q},\omega)\rangle.
\eeq 
Here $\mathbf{M}$ represents magnetic moments in global basis which can be obtained from 
 spin vectors in the global basis by multiplying with the $G-$tensors for various sublattice sites which are given in the appendix \ref{gmatrix} with $M_a^\alpha=G_a^{\alpha \beta} s^\beta_a.$  


%
%

\begin{figure}
\centering
\subfigure[$J_1=-1.0$; $J_2=-1.0$]{
\includegraphics[width=0.22\textwidth,height=3.0cm
]{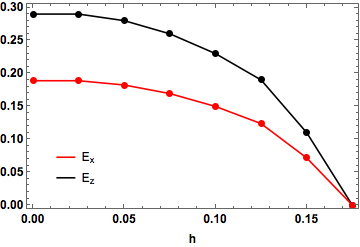}
}
\subfigure[$J_1=0.5$; $J_2=-2.0$]{
\includegraphics[width=0.22\textwidth,height=3.0cm]{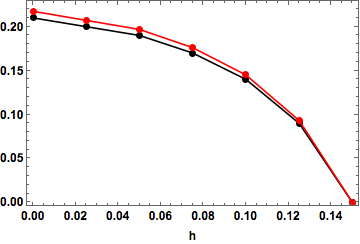}
}
\subfigure[$J_1=1.0$; $J_2=1.0$]{
\includegraphics[width=0.22\textwidth,height=3.0cm]{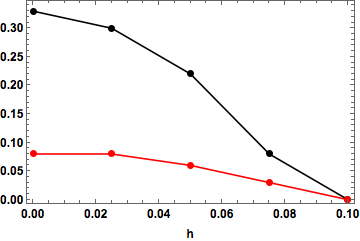}
}
\subfigure[$J_1=2.0$; $J_2=1.75$]{
\includegraphics[width=0.22\textwidth,height=3.0cm]{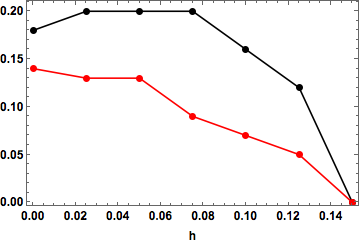}
}
\caption{Evolution of the independent mean field parameters  obtained by minimizing $E_{MF}$ in \eqn{eqn:EMF} and by setting the chemical potential self consistently satisfying the consraint in \eqn{eqn:mu}, due to the presence of a varying magnetic field $2h_Z[001]$ in global basis(the magnetic field term is added as in eqn. \ref{zeeman}). These figures shows that the uniform U(1) QSL phase can be destroyed by a global magnetic field $h_Z/|J_3| \sim \frac{c}{2}~ max(|\bar{E}_z|,|\bar{E}_x|)$, while $c\sim 1$. For all figures  $J_3=-1.0$; $J_4=0.0$.}
\label{fig:hz_pureqsl}
\end{figure}  
 
 The neutron scattering cross sections for representative parton band structures (corresponding to the same parameter values as fig. \ref{fig_bs_pure_qsl}) is given in fig. \ref{fig_sf_pure_qsl}. In a QSL, the dynamic spin structure factor probes the two-particle parton continuum as opposed to the sharp magnon excitation in a conventional magnetic ordered phase. Thus in a QSL, the dynamic spin structure factor is expected to be broad and diffused which is indeed one of the major motivations to understand the existing neutron scattering experiments of compounds such as Yb$_2$Ti$_2$O$_7$ and Er$_2$Sn$_2$O$_7$ starting from a QSL. The structure factors shown in eqn. \ref{fig_sf_pure_qsl} generically shows such broad and diffused scattering characteristic to the two-particle continuum. In case of parameter regimes where the parton band structure has a gap (\ref{fig_bs_pure_qsl} (b) and (c)) the corresponding 
 continuum in structure factor (\ref{fig_sf_pure_qsl} (b) and (c)) has a finite lower threshold of the order of $2\Delta_{\rm min}$, where $\Delta_{\rm min}$ is the minimum band gap for the partons. In case of gapless band structure (\ref{fig_bs_pure_qsl} (a) and (d)) the lower threshold goes down all the way to $\omega=0$ (\ref{fig_sf_pure_qsl}$(a)$ and $(d)$). However if the gapless regions represent points or small pockets, the spectral weight associated with them is small ({\it e.g.}, fig. \ref{fig_sf_pure_qsl}$(a)$). In this situation the the spectral weights from such a gapless QSL becomes practically indistinguishable from that of the QSL with a small gap.

\subsection{Effect of an external magnetic field on the uniform U(1) QSL phase}

We now turn to the effect of an external magnetic field on the uniform U(1) QSL. The magnetic field couples to the system through the usual Zeeman term
\begin{align}
H_{zeeman}=\sum_{i}h_\alpha g_{\alpha \beta}S_i^\beta
\label{zeeman}
\end{align}
where we have used the local basis and $g_{\alpha\beta}$ is the anisotropic $g$-factor given by \eqn{appen_g_local}. The spin is then written in terms of the fermionic partons using \eqn{eq_sploc} and self-consistent mean field solutions are obtained as before by adding the above Zeeman term to \eqn{eqn:Echi_hamiltonian_1}. In this calculations we have chosen the magnetic field to be in the $[001]$ direction.

\begin{figure}
\centering
\subfigure[$h_z=0.0$]{
\includegraphics[width=0.45\linewidth,height=3.0cm]{bs_1_1.png}
}
\subfigure[$h_z=0.075$]{
\includegraphics[width=0.45\linewidth,height=3.0cm]{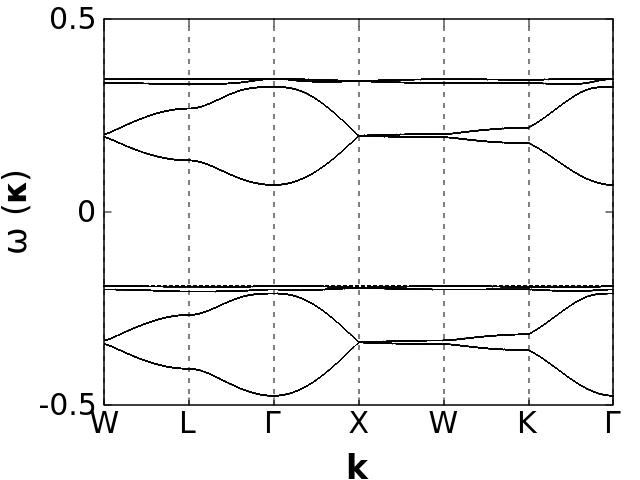}
}
\caption{Representative fermionic parton band structure in the presence of a global magnetic field $2h_Z[001]$, with $h_Z=0.075$. The parameters for this plot are $J_1=1.0$, $J_2=1.0$, $J_3=-1.0$, $J_4=0.0$. The dotted line in the  figure denotes the chemical potential.}
\label{qsl_hz}
\end{figure}

\begin{figure}
\centering
\subfigure[$h_z=0.0$]{
\includegraphics[width=0.45\linewidth,height=3.0cm]{sf_1_1_hd.png}
}
\subfigure[$h_z=0.075$]{
\includegraphics[width=0.45\linewidth,height=3.0cm]{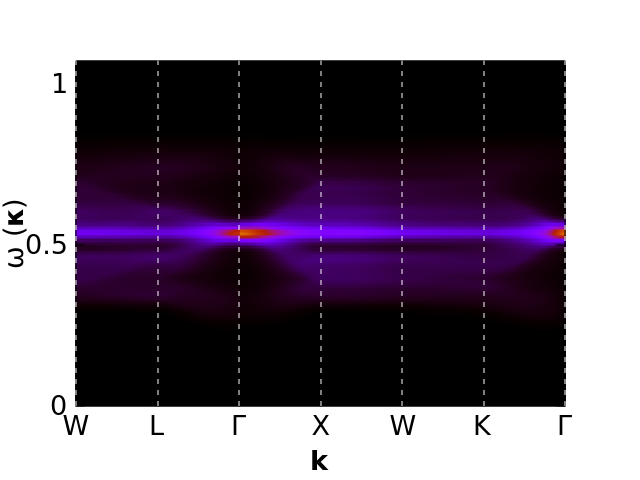}
}
\caption{The dynamical structure factor corresponding to the band structure in fig. \ref{qsl_hz}, in the presence of a global magnetic field $2h_Z[001]$, with $h_Z=0.075$. The parameters for this plot are $J_1=1.0$, $J_2=1.0$, $J_3=-1.0$, $J_4=0.0$. }
\label{fig_SF_QSL_h}
\end{figure}  
\begin{table*}[!htb]
\begin{tabular}{ccc}
     order~ & 
    $local$ ~ & 
    $global$~\\
\hline
\hline

$\mathbf{m}_E^{(1)}$ & $\frac{1}{2}(S_1^x+S_2^x +  S_3^x+S_4^x)$ & \hspace{-20mm} $\frac{-2 s_1^x+s_1^y+s_1^z+2 s_2^x-s_2^y+s_2^z+2 s_3^x+s_3^y-s_3^z-2 s_4^x-s_4^y-s_4^z}{2\sqrt{6}}$ 
  \vspace{2mm}\\
  $\mathbf{m}_E^{(2)}$ & $\frac{1}{2}( S_1^y + S_2^y +S_3^y+S_4^y)$ & $\frac{-s_1^y+s_1^z+s_2^y+s_2^z-s_3^y-s_3^z+s_4^y-s_4^z}{2\sqrt{2}}$\vspace{2mm} \\
$\mathbf{m}_{\text{T}_{1A'}}^{(3)}$ &\begin{tabular}{@{}c@{}}$\cos(t)\left(\frac{S_1^x+\sqrt{3}S_1^y+\sqrt{2}S_1^z + S_2^x+\sqrt{3}S_2^y+\sqrt{2}S_2^z-S_3^x-\sqrt{3}S_3^y-\sqrt{2}S_3^z-S_4^x-\sqrt{3}S_4^y-\sqrt{2}S_4^z}{2\sqrt{6}}\right)$\\$- \sin(t)
   \left(\frac{S_1^x+\sqrt{3}S_1^y+-2\sqrt{2}S_1^z + S_2^x+\sqrt{3}S_2^y-2\sqrt{2}S_2^z-S_3^x+\sqrt{3}S_3^y+2\sqrt{2}S_3^z-S_4^x-\sqrt{3}S_4^y+2\sqrt{2}S_4^z}{4\sqrt{3}}
\right)$ \end{tabular}&\begin{tabular}{@{}c@{}}$\frac{\cos (t)(s_1^z+s_2^z+s_3^z+s_4^z)}{2}$\\$-\frac{\sin(t)(-s_1^x-s_1^y+s_2^x+s_2^y-s_3^x+s_3^y+s_4^x-s_4^y)}{2\sqrt{2}}$ \end{tabular}\vspace{2mm}\\
$\mathbf{m}_{T_2}^{(3)}$ & $\frac{\sqrt{3}S_1^x - S_1^y+\sqrt{3}S_2^x - S_2^y - \sqrt{3} S_3^x+S_3^y-\sqrt{3}S_4^x+S_4^y}{4}$ & $\frac{-s_1^x+s_1^y+s_2^x-s_2^y-s_3^x-s_3^y+s_4^x+s_4^y}{2\sqrt{2}}$ \vspace{2mm}\\
\hline
\end{tabular}
\caption{The magnetic order parameters\cite{PhysRevB.95.094422} as chosen in our calculation (see eqns. \ref{eq_mag_1}, \ref{eq_mag_2} and \ref{eq_mag_3}).}
\label{m_order_params}
\end{table*}

 Fig. \ref{fig:hz_pureqsl} shows the evolution of self consistently determined independent QSL mean field parameters with an external magnetic field $h_Z$, for the four representative points in the phase diagram (fig. \ref{fig_tbi}). With the increasing field,  the uniform U(1) QSL phase gradually destroyed. We find that the QSL phase is completely destroyed with a small relatively magnetic field $h_Z/|J_3| \sim \frac{c}{2}~ max(|E_z|,|E_x|)$, while $c$ is close to 1. Simultaneously the magnetisation develops (not shown). For the partons the decrease in the QSL parameters is reflected in the reduction of their band width of individual bands as seen in fig. \ref{qsl_hz}. This directly translates into the narrowing of the region of finite intensity of the dynamic structure-factor as shown in fig. \ref{fig_SF_QSL_h} and gradual disappearance of the diffuse two-particle continuum. However, since our current calculation does not keep into account of the magnon excitations in a finite magnetically  polarized state, the resultant sharp excitation forms a flat structure of intensity that is particularly prominent in fig. \ref{fig_SF_QSL_h}$(b)$ and is reminiscent of sharp magnon modes at high magnetic fields in Yb$_2$Ti$_2$O$_7$.
 
 This completes our discussion of the uniform U(1) QSL phase with fermionic partons and we now turn to investigate their magnetic instability.


\section{Competition between the uniform U(1) QSL and magnetic ordered phases}
\label{section:magnetic_order}

To understand the nature of the magnetic instability of the above QSL, we shall now add the magnetic ordering channels in addition to the QSL channels (eq. \ref{eq_qsl_mf}). Unlike the QSL, however, the magnetic channels are gauge invariant and finite expectation values for these channels leads to spontaneous broken symmetry.  In particular $\langle {\bf S}\rangle\neq 0$ breaks time reversal as well as lattice symmetries.  We shall focus on the ${\bf q}=0$ magnetic ground states obtained in the classical limit of the the Hamiltonian in Eq. \ref{eqn:globalH} that have been found to be competitive in the regime of interest. \cite{PhysRevB.95.094422} For the ${\bf q}=0$ phases, the classical order parameters that appear are  the ones which transform as various irreducible representations of the symmetry group $\mathcal{T}_d.$ In particular, for the above Hamiltonian on a tetrahedron, the magnetic order parameters that contribute are a singlet $\mathbf{m}_{A_2},$ three triplet order parameters $\mathbf{m}_{T_2}, \mathbf{m}_{T_{1A'}},\mathbf{m}_{T_{1B'}},$ and a doublet  $\mathbf{m}_E$ transforming under $ A_2,\, T_2,\, T_1,\,T_1,\,\text{ and } E $ irreducible representations of $\mathcal{T}_d$ respectively. \cite{PhysRevB.95.094422} The detailed expressions of the $\mathbf{m}$'s in terms of the spins are given in Ref. \onlinecite{PhysRevB.95.094422} while we list the relevant ones in Table \ref{m_order_params}.\\
The Hamiltonian (\eqn{eqn:globalH}) can be exactly represented in terms of these order parameters channels as:
\bea\mylabel{eqn:H_MO}
H^{mag}&=&\frac{1}{2}\sum_{\boxtimes}a_{A_2} \textbf{m}_{A_2}^2+ a_{E}\textbf{m}_E^2 +a_{T_2} \textbf{m}_{T_2}^2 ~~~~~~~~~~~~~~~~~~\nonumber\\
&~&~~~~~~+a_{\text{T}_{1A'}}\textbf{m}_{\text{T}_{1A'}}^2 + a_{\text{T}_{1B'}}\textbf{m}_{\text{T}_{1B'}}^2.~~~~~~~~~~~~~~~~~
\eea

In \eqn{eqn:H_MO}, the sum is over the tetrahedron and $a_{A_2},~a_E,~a_{T_2},~a_{T_{1A'}},~a_{T_{1B'}}$ are coupling constants which are given by linear combinations of $J_\alpha$ (Table \ref{a_coeff} in Appendix \ref{appen_coupling}). The classical phase diagram for the above Hamiltonian(\eqn{eqn:H_MO}) is worked out in detail in Ref. \onlinecite{PhysRevB.95.094422}. Below we give a brief overview of these phases for completeness. Given the positive quadratic forms, the boundaries of the classical phases can be estimated by comparing the coefficients of \eqn{eqn:H_MO}.

In the parameter regime of relevance that we consider, $a_E, a_{T_2}$ and $a_{\text{T}_{1A'}}$ have the minimum values, so we only consider the corresponding  order parameters $\mathbf{m}_E, \mathbf{m}_{T_2}, \mathbf{m}_{T_{1A'}}.$  Without loss of generality, we choose the directions of mean field order parameters to be 

\begin{align}
\bar{\mathbf{m}}_E= \bar{m}_E \left(
\begin{array}{c}
  \cos(\pi/6) \\
\sin(\pi/6)
\end{array}
\right)
\equiv \left(
\begin{array}{c}
 \langle\textbf{m}_E^{(1)}\rangle \\
 \langle\textbf{m}_E^{(2)}\rangle
\end{array}
\right),
\label{eq_mag_1}
\end{align} 

\begin{align}
\bar{\mathbf{m}}_{T_2}= \langle\textbf{m}_{T_2}^{(3)}\rangle \left(
\begin{array}{c}
 0 \\
 0\\
 1
\end{array}
\right),
\label{eq_mag_2}
\end{align}
 and
 \begin{align}
 \bar{\mathbf{m}}_{T_{1A'}}=\langle \textbf{m}_{T_{1A'}}^{(3)}\rangle \left(
\begin{array}{c}
 0 \\
 0\\
 1
\end{array}
\right)
\label{eq_mag_3}
\end{align}
 and solve for the ground state of eqn. \eqn{eqn:H_MO}. Here $\mathbf{m}_X^{(\alpha)} $ refers to the $\alpha^{\text{th}}$ component of $\mathbf{m}_X$ (Table \ref{m_order_params}).

The phase boundaries between these classical magnetically ordered  phases are shown by  dashed lines in fig. \ref{fig_pd}. In the region between $a_{T_{1A'}}=a_{T_2}$ and $a_{T_{1A'}}=a_{E}$, where $J_1,J_2(<0)$, a ferromagnetic phase characterised by $\bar{\bf m}_{T_{1A'}}\neq 0$ is stabilized. The ground state is in fact a non-colinear ferromagnet (splayed ferromagnet (SF)) where the spin configurations in a tetrahedron are given by $\textbf{S}_a \equiv \{\pm\frac{\sin t}{\sqrt{2}},\mp\frac{\sin t}{\sqrt{2}},\frac{\cos t}{\sqrt{2}}\}$, $t$ being the canting angle between spins and the $[100]$ axis in the ferromagnetic ground state\cite{PhysRevB.95.094422}. 

The region between the $a_{E}=a_{T_2}$ and $a_{T_{1A'}}=a_{E}$, with antiferromagnetic XY interactions $J_1>0$; the ground state, characterised by $\bar{\textbf{m}}_E\neq 0$, belongs to a one-dimensional manifold of states which transforms under the $E$ irreducible representation of the $\mathcal{T}_d$ group. These states are charecterised as classical spins lying in the XY plane normal to the local $[111]$ axis on each site, with spin configurations 
$\textbf{S}_a \equiv \{(\pm \frac{\sin\theta_E}{\sqrt{2}},\pm \frac{\sin\theta_E}{\sqrt{2}}, \frac{\cos\theta_E}{\sqrt{2}}),(\pm \frac{\sin\theta_E}{\sqrt{2}},\mp \frac{\sin \theta_E}{\sqrt{2}}, -\frac{\cos \theta_E)}{\sqrt{2}})\}$. It is understood \cite{PhysRevB.95.094422} that for the parametric regime near the ferromagnetic phase boundary $a_{T_{1A'}}=a_{E}$, fluctuations select a coplanar antiferromagnetic phase described as 
$\theta_E=\frac{(2m+1)\pi}{6},~ m=0,1,...,5$. However in other regions fluctuations favor a non-coplanar antiferromagnetic phase described as $\theta_E=\frac{(n)\pi}{3},m=0,1,...,5$. For the $m_E$ phase (also referred to as $\Psi_3$ phase), we choose $\theta_E = \pi/6$ in our analysis.  

Finally, the region between the $a_{E}=a_{T_2}$ and $a_{T_{1A'}}=a_{T_2}$,  the ground state is characterized by $\bar{\bf m}_{T_2}$ (the Palmer-Chalker phase, $PC$), as a helical spin configuration in a common $[100]$ plane and the spin configuration reads as $\textbf{S}_a \equiv 
\{(\pm \frac{1}{\sqrt{2}},\pm \frac{1}{\sqrt{2}}, 0),(\pm \frac{1}{\sqrt{2}},\mp \frac{1}{\sqrt{2}}, 0)\}$\citep{PhysRevB.95.094422}.

\subsection{Magnetic instability to the QSL}

With these magnetic order we now try to understand the instability of the uniform U(1) QSL to one these phases within a parton mean-field theory augmented with the magnetic channels as
\begin{align}
\tilde{H}_{MF}=H_{MF}+\alpha H_{MFT}^{mag}
\label{eq_hybrid}
\end{align}
where $H_{MF}$ is the pure U(1) parton mean-field Hamiltonian (eqn. \ref{eqn:HMF}) and $H_{MFT}^{mag}$ is the Curie-Weiss mean field Hamiltonian for the magnetic order under consideration where the spin operators are replaced by the parton operators using eqn. \ref{eq_sploc}. To understand the competition and coexistence of magnetic phases and QSL phase, we take the total mean field Hamiltonian including both types of mean field channels. Due to the difference of renormalisation\cite{chern2018magnetic} of the QSL mean field channels and the magnetic moments the relative weightage between the QSL decoupling and magnetic decoupling is in general not equal and is parametrised by $\alpha$ in eqn. \ref{eq_hybrid}. Being a renormalisation effect, the value of $\alpha$ cannot be determined within the mean field theory and hence similar to Ref. \onlinecite{chern2018magnetic,PhysRevB.69.035111}, we shall take it as a variational parameter to sketch out the possible interplay between the QSL and the magnetic order. For example, if we take the QSL and magnetic Hamiltonians with equal weight, the classical magnetic phases always win. On decreasing $\alpha$ from $1$, the QSL becomes competitive and in the rest of this work we shall present results for $\alpha=1/4$.

\begin{figure}
\includegraphics[scale=0.65]{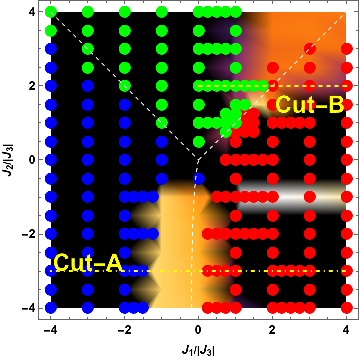}
\caption[unitcell]{The phase diagram of a pyrochlore magnet described by the Hamiltonian \eqn{eqn:globalH}, obtained by minimizing \eqn{eqn:EMF_tot} with respect to $\bar{E}^x$, $\bar{E}^z$, $\bar{m}_{\text{E}}$, $\bar{m}_{\text{T}_{1A'}}$ and $\bar{m}_{\text{T}_{2}}$. The data points shown in blue has a magnetically ordered phase with non-collinear ferromagnetic order(splayed ferromagnet, $SF$) characterized by order parameter $\bar{m}_{\text{T}_{1A'}}$ as described in section~\ref{section:magnetic_order}; the  green data points represent a magnetically ordered phase with Palmer-Chalker order charecterised by order parameter $\bar{m}_{\text{T}_{2}}$ and the region shown in red is a planar antiferromagnet phase($E$) charecterised by the order parameter  $\bar{m}_{\text{E}}$. The regions of the phase diagram where the U(1) QSL order parameters ($\bar{E}^x$, $\bar{E}^z$)are non-zero, is marked with shades of yellow with intensity proportional to $max(|\bar{E}_z|,|\bar{E}_x|)$ , and the data is extrapolated for a continuous plot for these regions. The phase diagram obtained from our calculation shows regions of pure U(1) QSL phase and regions of coexisting magnetic order and U(1) QSL phase. For all points $J_3=-1.0$; $J_4=0.0$ (and $\alpha=0.25$). The parameters along the two cuts $A$ and $B$ are given in fig. \ref{fig_cut}.}
\label{fig:pd_dot}
\end{figure}

\begin{figure}
\centering
\includegraphics[scale=0.65]{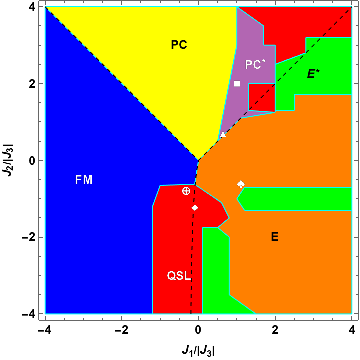}
\caption{The extrapolated phase diagram obtained from the data points of fig. \ref{fig:pd_dot}. The region shown in blue has a magnetically ordered phase with non-colinear ferromagnetic order(splayed ferromagnet, $FM$) charecterised by order parameter $\bar{m}_{\text{T}_{1A'}}$ (section~\ref{section:magnetic_order}); the region shown in yellow is a magnetically ordered phase with Palmer-Chalker ($PC$) order charecterised by order parameter $\bar{m}_{\text{T}_{2}}$ and the region shown in orange is a planar antiferromagnet phase charecterised by the order parameter  $\bar{m}_{\text{E}}$. On the other hand, the QSL region is in red. There are two fractionalised magnetically ordered phases--(1) the fractionalised PC phase in purple ($PC^*$), and (2) the fractionalised XY phase in green ($E^*$). For all points $J_3=-1.0$; $J_4=0.0$ and $\alpha=0.25$. The following symbols denote the position of the interesting materials on this hyper-plane : (1) $(\boldsymbol{\oplus})$ for Yb$_2$Ti$_2$O$_7$ from Ref. \onlinecite{ross2011quantum}, (2) $(\boldsymbol{\blacklozenge}) $  for Yb$_2$Ti$_2$O$_7$ from Ref. \onlinecite{PhysRevLett.119.057203},  (3) $(\boldsymbol{\spadesuit})$ for Er$_2$Ti$_2$O$_7$, (4)  $(\boldsymbol{\blacksquare})$ for  Er$_2$Sn$_2$O$_7$, and (5) $(\boldsymbol{\blacktriangle})$ for Er$_2$Pt$_2$O$_7$.}
\label{fig_pd}
\end{figure}

The resultant phase diagram is shown in figs. \ref{fig:pd_dot} and \ref{fig_pd} which gives an indication for the competition between the magnetic orders and the QSL while the quantitative details of the phase diagram needs to be worked out with more sophisticated numerical techniques. In the rest of this section we provide details of the obtained phase diagram and also discuss its possible implications to the physics of rare-earth pyrochlores. 

The total mean-field Hamiltonian $\tilde{H}_{MF}$ is now given by:
\bea\mylabel{HMF_MO_terms}
\tilde{H}_{MF}&=&\frac{3N}{16}   \big[ 2(J_{zz}+4J_{\pm})|\bar{E}^z|^2 -8(J_{zz} -4J_{\pm\pm})|\bar{E}^x|^2 \big] \nonumber \\
&+&\frac{3N}{16}  \big[ 16 J_{z\pm}( \bar{E}_{12}^{z\dagger} \bar{E}_{12}^{x} + \bar{E}_{12}^{x\dagger} \bar{E}_{12}^{z})] \nonumber \\
&-&\alpha \frac{N}{4}(a_E ~\bar{\mathbf{m}}_{\text{E}}^2 +a_{\text{T1A}'}~\bar{\mathbf{m}}_{\text{T}_{1A'}}^2
   +a_{\text{T2}} ~\bar{\mathbf{m}}_{\text{T}_{2}}^2  )\nonumber\\
&+& \sum_{\bf{k}\in BZ} f_a^{\sigma_1\dagger}(\textbf{k}) \tilde{H}^{ab}_{\sigma_1 \sigma_2}(\textbf{k}) f_b^{\sigma_2}(\textbf{k}),
\eea
where 
$\tilde{\textbf{H}}$ now has the magnetic terms on the diagonal blocks.  The elements of the diagonal magnetic blocks are given in the Appendix \ref{appen_MFT_details}.

\begin{table}
\begin{tabular}{ | c | c | c | c | c | c | c | c |}
\hline
     $J_1$~ & 
     $J_1$~ & 
    $\bar{E}_z$~ & 
    $\bar{E}_x$~ &
    $\bar{m}_{\text{T}_{1A'}}$~ & 
    $\bar{m}_{\text{E}}$~ & 
    $\bar{m}_{\text{T}_{2}}$~ & 
    $\tilde{E}_{MF}$ \\
\hline
\hline
$2.0$ & $1.5$ & $-0.2$ & $-0.1$  & $0.0$  & $-0.2$ & $-0.1$  & $-0.238$ \\
\hline
$0.5$ & $-3.0$ & $-0.18$ & $-0.16$  & $0.0$  & $-0.33$ & $0.0$  & $-0.332$\\
\hline
$1.75$ & $2.0$ & $-0.06$ & $-0.17$  & $0.0$  & $0.0$ & $-0.4$  & $-0.223$\\
\hline
\end{tabular}
\caption{ Mean field parameters values in the local basis($\bar{E}^x$, $\bar{E}^z$, $\bar{m}_{\text{E}}$, $\bar{m}_{\text{T}_{1A'}}$ and $\bar{m}_{\text{T}_{2}}$, in Col. 3, 4, 5, 6 and 7 respectively) obtained by minimizing the mean-field energy per site as in \eqn{eqn:EMF_tot}. The mean-field energy minimum values per site ($\tilde{E}_{MF}$) are given in Col. 8. Results are shown for the parameters values $J_1$ and $J_2$ as given in Col.1 and 2 respectively, keeping the parameters $J_3=-1.0, J_4=0.0$ fixed. The points in this table are representative points of coexisting U(1) QSL quantum order and classical magnetic order and the points are marked as $a,b,c$ in fig. \ref{fig_pd}.}
\label{QSL_mo_params}
\end{table}
The mean-field ground-state of this augmented Hamiltonian of combined parton mean field channels and magnetic channels (\eqn{HMF_MO_terms}) is obtained like before (see sub-section \ref{pure_qsl_minimization}) by filling in the single particle energies for the band-structure where the mean-field parameters $\bar{E}^x$, $\bar{E}^z$, $\bar{m}_{\text{E}}$, $\bar{m}_{\text{T}_{1A'}}$ and $\bar{m}_{\text{T}_{2}}$(in local basis, say) are determined self-consistently with the constraint of one parton per site implemented on average using a chemical potential $\mu$, which is also calculated self-consistently. This is done by minimising the mean-field energy per site $\tilde{E}_{MF}$ as given by \eqn{eqn:EMF_tot} in the local basis, the $\mu$ is determined at each step of the minimization by solving \eqn{eqn:mu_tot} for the chemical potential (details in Appendix \ref{appen_mft_min}).

\begin{figure}
\subfigure[]{
\includegraphics[width=0.225\textwidth,height=3.5cm]{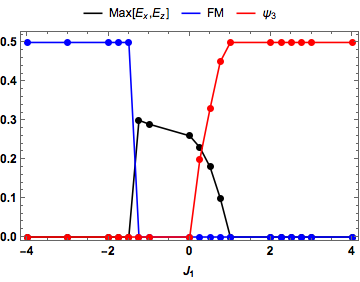}
}
\subfigure[]{
\includegraphics[width=0.225\textwidth,height=3.5cm
]{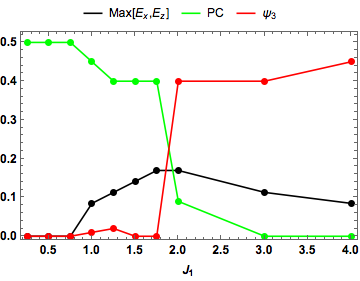}
}
\caption{The variations of various QSL and magnetic mean-field parameters along Cut A and Cut B of fig. \ref{fig:pd_dot}.}
\label{fig_cut}
\end{figure}

\bea\mylabel{eqn:EMF_tot}
\tilde{E}_{MF}&=& \frac{3}{16} \big[ 2(J_{zz}+4J_{\pm})|\bar{E}^z|^2 +8(-J_{zz} +4J_{\pm\pm})|\bar{E}^x|^2 \big]\nonumber\\
&&+\frac{3}{16}\big[ 16 J_{z\pm}( \bar{E}_{12}^{z\dagger} \bar{E}_{12}^{x} + \bar{E}_{12}^{x\dagger} \bar{E}_{12}^{z})]\\ \nonumber
&&- \frac{\alpha}{4}(a_E ~\bar{\mathbf{m}}_{\text{E}}^2 +a_{\text{T1A}'}~\bar{\mathbf{m}}_{\text{T}_{1A'}}^2
   +a_{\text{T2}} ~\bar{\mathbf{m}}_{\text{T}_{2}}^2  )\\ \nonumber
&&+ \frac{1}{4 N_{UC}} \sum_{{\bf k} \in BZ} \sum_{i=1}^{N_{band}}  \tilde{\Lambda}_i(\textbf{k}) \Theta\big[\mu,\tilde{\Lambda}_i(\textbf{k}) \big].
\eea
\beq\mylabel{eqn:mu_tot}
\frac{2}{N_{band}N_{UC}}\sum_{{\bf k} \in BZ} ~\sum_{i=1}^{N_{band}}~\Theta\big[\mu,\tilde{\Lambda}_i(\textbf{k}) \big]=1.
\eeq
where $\tilde{\Lambda}_i$~s are the eigenvalues of $\tilde{H}^{ab}_{\sigma_1 \sigma_2}$, rest of the notations are same as is \eqn{eqn:EMF} and \eqn{eqn:mu}. The parametric values of the independent mean field parameters in the local basis and corresponding mean field energy per site is given in TABLE \ref{QSL_mo_params} for some representative points in the phase diagram.

 The above self consistent parton mean field  calculations, augmented with magnetic channels gives rise to a very rich phase diagram. The different non zero QSL and magnetic order parameters are shown in fig. \ref{fig:pd_dot} while the dependence of various mean field parameters along the two cuts of fig. \ref{fig:pd_dot} are shown in fig. \ref{fig_cut}. The expectation that the QSL becomes competitive near the boundaries of the magnetic phases is indeed realised. In addition, the planar antiferromagnet and the Palmer-Chalker magnetic orders can coexist with the QSL as seen both from fig. \ref{fig:pd_dot} and \ref{fig_cut}. Such magnetically ordered states which allows fractionalisation are indeed attractive candidates for rare-earth pyrochlores where experiments seem to find signatures of both. We however note that in our calculations, no co-existing region between the QSL and the ferromagnetic phase was observed. We shall return to this point when we discuss the relevance of our calculations to candidate materials.  The interpolation of the data points produce the phase diagram shown in fig. \ref{fig_pd} where we have pointed out the different phases. Clearly the proximity of the QSL to the classical phase boundaries bears out our expectation about significant quantum fluctuations leading to long range entangled phases in regions where {\it classical} orders are fragile. In the figure we have also plotted the projection of the position of the interesting materials on this hyper-plane based on experimental estimates of the magnetic exchange couplings (see Table \ref{tab_coupl}). These are Yb$_2$Ti$_2$O$_7$, Er$_2$Ti$_2$O$_7$,   Er$_2$Sn$_2$O$_7$ and  Er$_2$Pt$_2$O$_7$. We now turn to the discussion of the relevance of our results to these materials.
 
\subsection{Structure factor}

\begin{figure}
\centering
\subfigure[$J_1=2.0$, $J_2=1.5$]{
\includegraphics[width=0.3\linewidth,height=2.5cm]{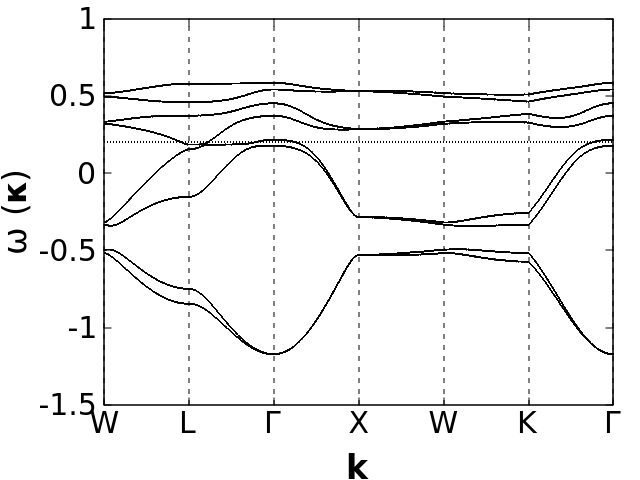}
} 
\subfigure[$J_1=0.5$, $J_2=-3.0$]{
\includegraphics[width=0.3\linewidth,height=2.5cm]{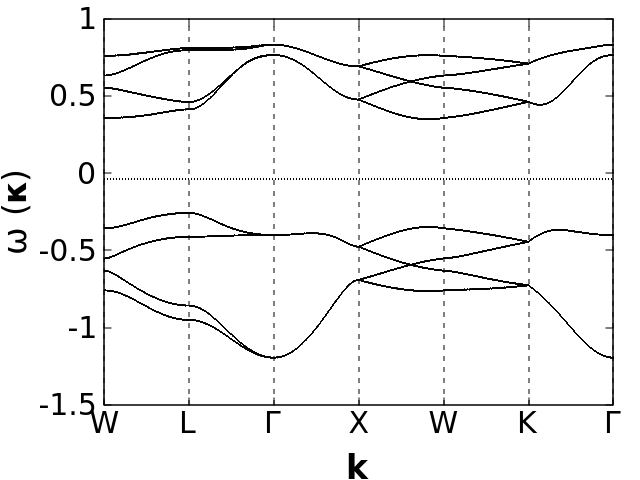}
}
\subfigure[$J_1=1.75$, $J_2=2.0$]{
\includegraphics[width=0.3\linewidth,height=2.5cm]{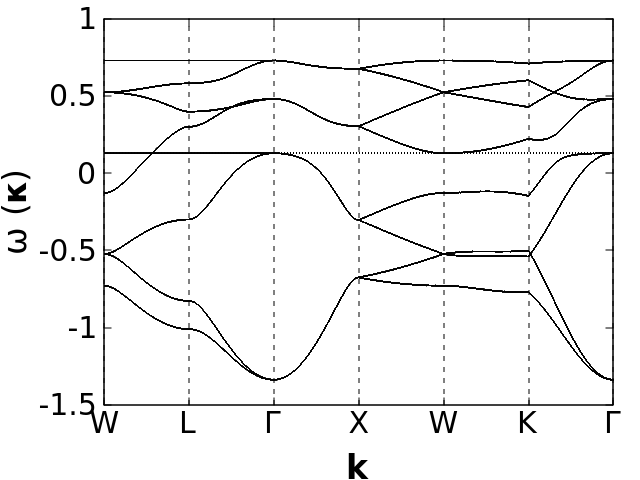}
}
\caption{Representative fermionic parton band structures for the points $a,b,c$ of fig. \ref{fig_pd}, which has coexisting magnetic order and U(1) QSL phase. The two-fold degeneracy of the bands are lifted due the broken TR invariance by the magnetic order. The independent mean field parameters are obtained by minimizing $\tilde{E}_{MF}$ in \eqn{eqn:EMF_tot} and by setting the chemical potential self consistently satisfying the constraint in \eqn{eqn:mu_tot}, the parameters are listed in Table.~\ref{QSL_mo_params}. The dotted line in each figures denotes the chemical potential. Fig. \ref{fig:sf_m2}$(a)$ shows Fermi and hole pockets and predicts gapless excitations. Fig. \ref{fig:sf_m2}$(b)$ predicts gapped excitations and fig. \ref{fig:sf_m2}$(c)$ predicts gapless excitations. For all figures  $J_3=-1.0$; $J_4=0.0$. }
\label{fig:sf_m2}
\end{figure}

\begin{figure}
\centering
\subfigure[$J_1=2.0$; $J_2=1.5$]{
\includegraphics[scale=0.32]{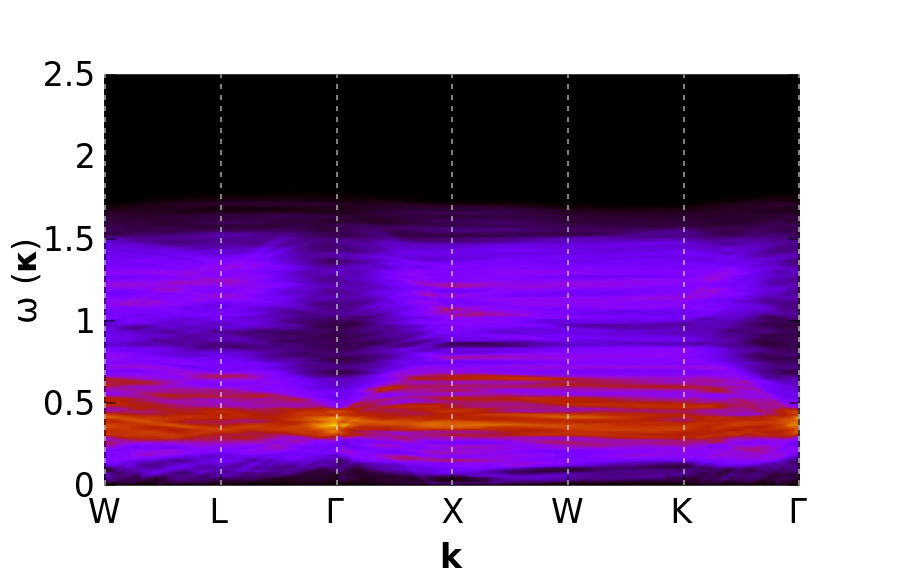}
}
\subfigure[$J_1=0.5$; $J_2=-3.0$]{
\includegraphics[scale=0.36]{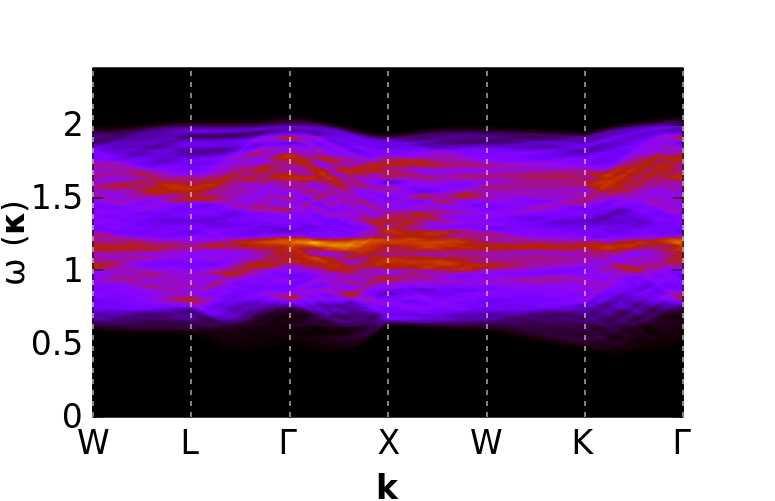}
}
\subfigure[$J_1=1.75$; $J_2=2.0$]{
\includegraphics[scale=0.39]{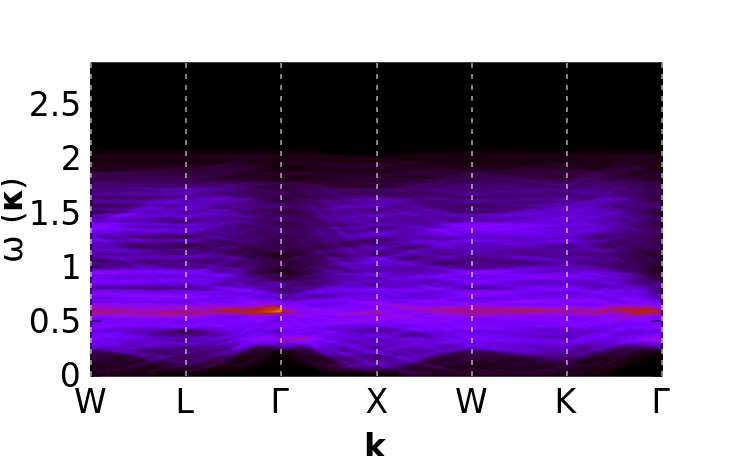}
}
\caption{Representative dynamical structure factors corresponding to the band structures in fig. \ref{fig:sf_m2}. Fig. \ref{fig:sf_m2_3}$(a)$ and \ref{fig:sf_m2_3}$(cw)$ shows gapless dispersive band structure due to the gaplss band structure of fig. \ref{fig:sf_m2}$(a)$ and \ref{fig:sf_m2}$(c)$. Fig. \ref{fig:sf_m2_3}$(b)$ shows a gapped excitations due to the gapped band structure of fig. \ref{fig:sf_m2}$(b)$. For all figures  $J_3=-1.0$; $J_4=0.0$. }
\label{fig:sf_m2_3}
\end{figure}

 For calculating the structure factor in presence of the magnetic orders, we exclusively focus on the regions where the magnetic orders coexist with the QSL as these are the more non-trivial cases as far as the present calculations are concerned. The three representative parton band-structures are shown in fig. \ref{fig:sf_m2}. The band are no longer doubly degenerate as time reversal symmetry is broken in presence of the magnetic order. In the first two cases the partons are gapped while in the third case it is gapless. The corresponding structure factors are shown in fig. \ref{fig:sf_m2_3}. The diffuse continuum is quite prominent in all the cases. Neutron scattering in these states therefore will see quasi-elastic continuum in addition to the magnetic Bragg peaks. Due to the decrease in the magnitude of the QSL parameters in this regime, the band-width of the parton bands decreases which results in squeezing of region of non-zero intensity of the dynamic structure factor.

\subsection{Effect of the magnetic field}

\begin{figure*}
\centering
\subfigure[$h_z=0.0$]{
\includegraphics[scale=0.21]{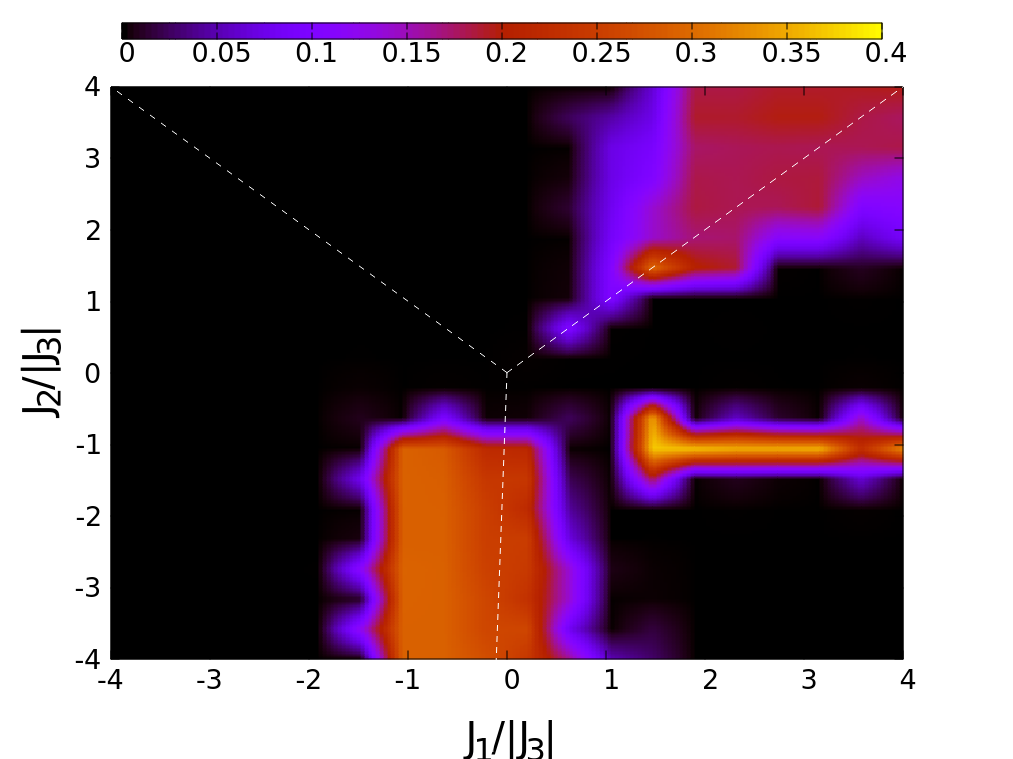}
}
\subfigure[$h_z=0.025$]{
\includegraphics[scale=0.21]{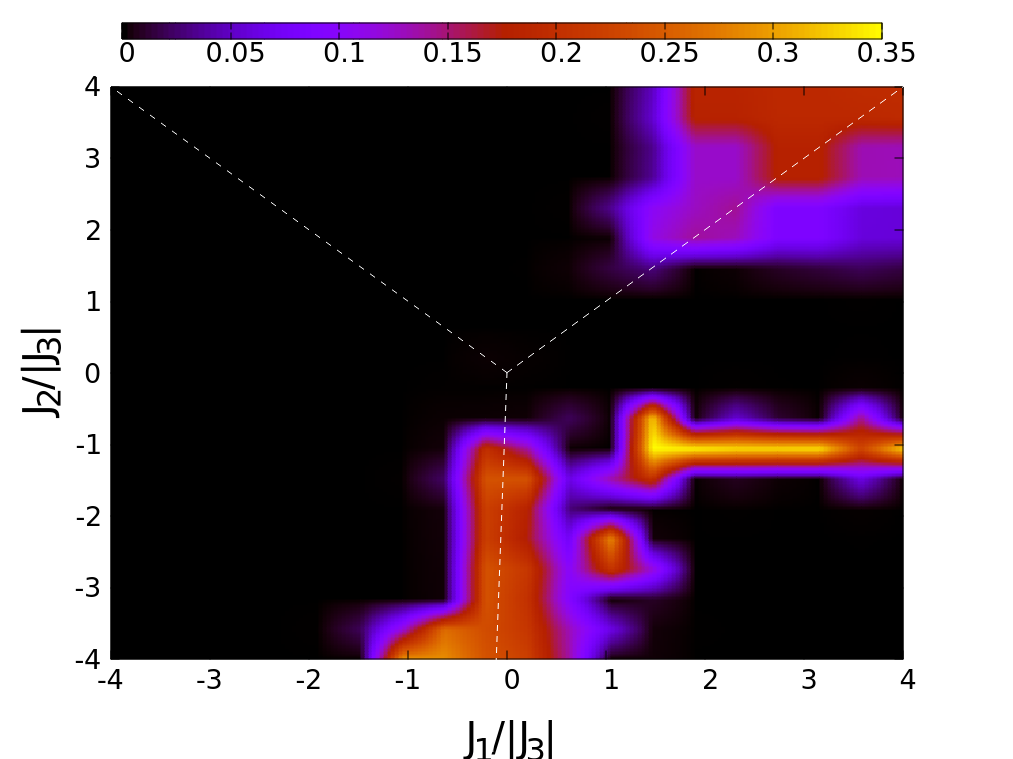}
}
\subfigure[$h_z=0.05$]{
\includegraphics[scale=0.21]{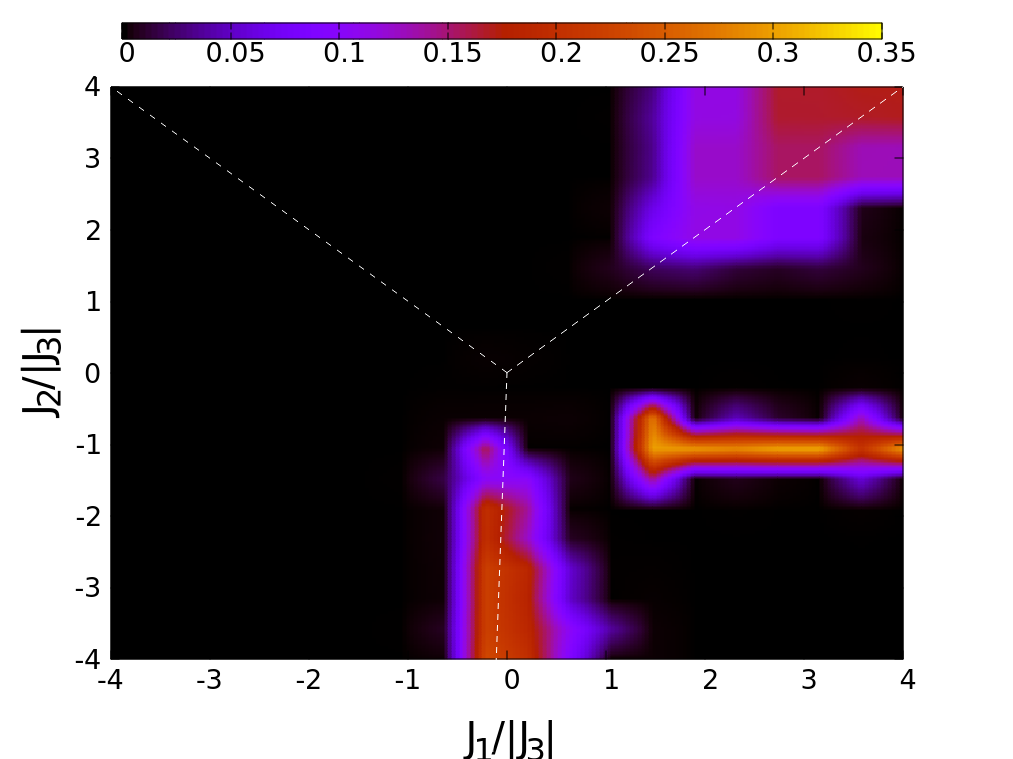}
}
\caption{Evolution of the U(1) QSL phase (not showing the magnetic order which is non zero here) boundary is shown with a magnetic field $2h_z[001]$. The parameters of this phase diagram is obtained by minimizing \eqn{eqn:EMF_tot} with additional global magnetic field, with respect to $\bar{E}^x$, $\bar{E}^z$, $\bar{m}_{\text{E}}$, $\bar{m}_{\text{T}_{1A'}}$ and $\bar{m}_{\text{T}_{2}}$. Our results shows that the U(1) QSL region shrinks with the external magnetic field. For all points $J_3=-1.0$; $J_4=0.0$ and $\alpha=0.25$. }
\label{fig_bs_pqsl}
\end{figure*}

Finally we turn to the effect of an external magnetic field on the competition between the magnetically ordered phases and the QSL. The Zeeman coupling is given by eqn. \ref{zeeman} with the magnetic field being in the $[001]$ direction. We consider the evolution of the QSL phases under this magnetic field and this is shown in fig. \ref{fig_bs_pqsl} The decreasing regime of the QSL shows its instability to the polarised state which wins over at high magnetic field. 

\begin{figure}
\centering
\subfigure[$h_z=0.0$]{
\includegraphics[width=0.45\linewidth,height=3.0cm]{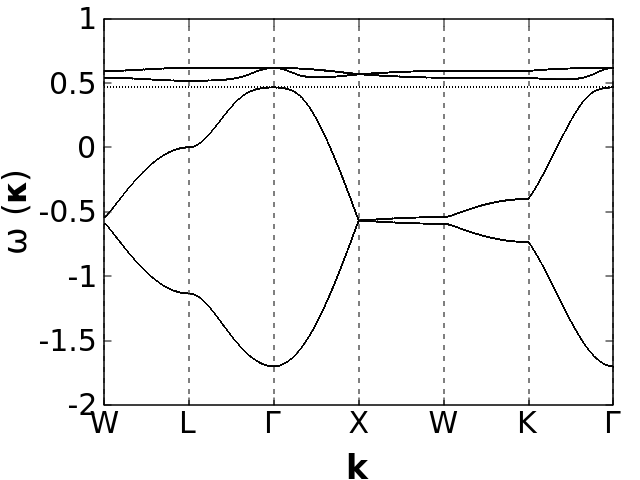}
}
\subfigure[$h_z=0.025$]{
\includegraphics[width=0.45\linewidth,height=3.0cm]{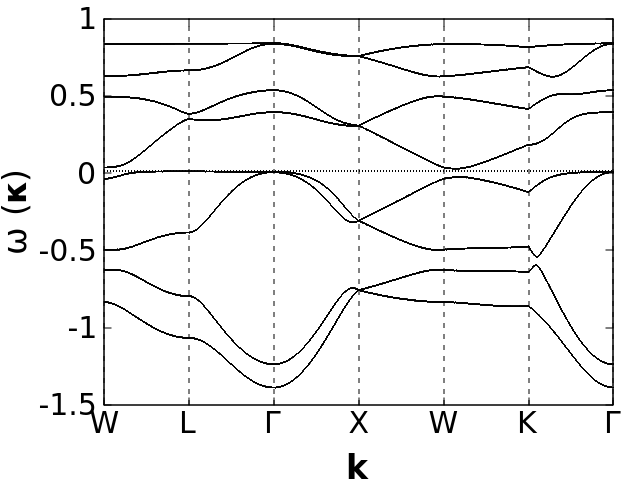}
}
\caption{Representative fermionic parton band structure in the presence of a global magnetic field $2h_Z[001]$, with $h_Z=0.025$. The parameters for this plot are $J_1=2.0$, $J_2=3.0$, $J_3=-1.0$, $J_4=0.0$; the mean field calculations shows a coexistence of U(1) QSL phase and planer antiferrogmagnetic order. The dotted line in the  figure denotes the chemical potential.}
\label{qsl_mo_hz}
\end{figure}

\begin{figure}
\centering
\subfigure[$h_z=0.0$]{
\includegraphics[width=0.45\linewidth,height=3.0cm]{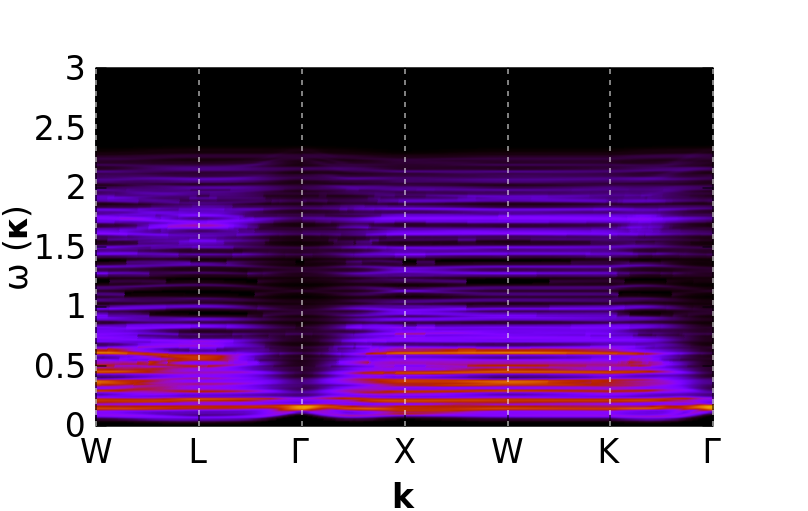}
}
\subfigure[$h_z=0.025$]{
\includegraphics[width=0.45\linewidth,height=3.0cm]{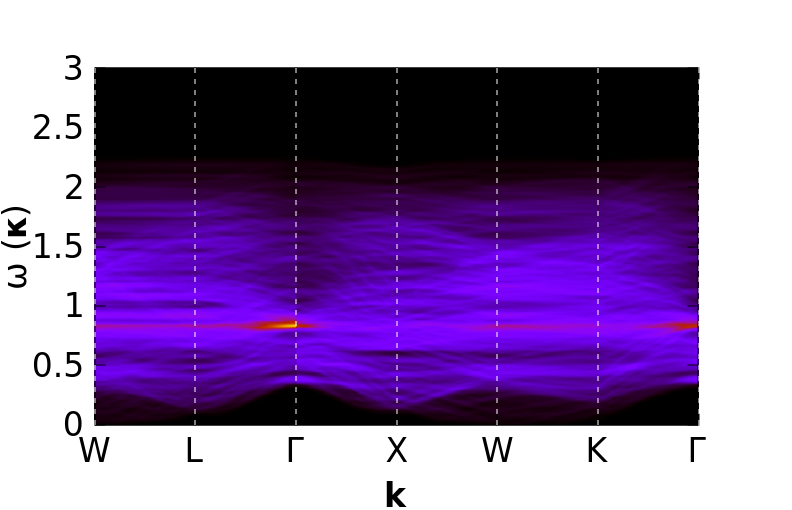}
}
\caption{The dynamical structure factor corresponding to the band structure in fig. \ref{qsl_mo_hz}, in the presence of a global magnetic field $2h_Z[001]$, with $h_Z=0.025$. The parameters for this plot are $J_1=2.0$, $J_2=3.0$, $J_3=-1.0$, $J_4=0.0$.}
\label{fig_sf_in_mo_h}
\end{figure}

The change in the parton band structure in at representative coexisting (QSL and magnetic order) point is shown in fig. \ref{qsl_mo_hz} and the corresponding changes in the structure factor is shown in fig. \ref{fig_sf_in_mo_h}. Clearly, the QSL parameters decreases in magnitude as the magnetic field is increased which leads to the development of magnetic polarization. This is reflected in both the parton band structure as well as the structure factor.

\section{Relevance to materials}
\mylabel{sec:Materials}
In regards to the relevance of our results to materials, we shall primarily focus on Yb$_2$Ti$_2$O$_7$ and Er$_2$Sn$_2$O$_7$ which are near the phase classical phase boundary as far as the present experimentally determined coupling constants are concerned. We shall also comment on the others briefly. 

\paragraph*{{\rm Yb$_2$Ti$_2$O$_7$ :}} The present calculations seems to suggest that {\rm Yb$_2$Ti$_2$O$_7$ is in a $U(1)$ QSL phase proximate to a slay ferromagnet (fig. \ref{fig_pd}). However, Ref. \onlinecite{chern2018magnetic} or \onlinecite{savary2012coulombic}, in the present case within parton mean field description, our calculations does not capture a fractionalised ferromagnetic phase where both fractionalised partons and ferromagnetic order coexist. In fig. \ref{fig:bd_ybto}, we plot the band structure for the self consistently derived mean-field solution with the coupling constants estimated from Ref. \onlinecite{PhysRevLett.119.057203} and \onlinecite{ross2011quantum} respectively. It is clear that for both the cases, the resultant time reversal invariant $U(1)$ QSL has a gapped band structure for the fermionic partons and the band structure has a non-trivial $Z_2$ topological invariant. We note that such a fractionalized topological insulators or the so-called $(E_{fT}M_f)_\theta$ was also suggested for Yb$_2$Ti$_2$O$_7$ in Ref. \onlinecite{PhysRevX.6.011034}. The relevance of this novel symmetry enriched topological phase, however, is not clear.   Attempts to understand the resultant gapless surface states through thermal conductivity\cite{tokiwa2016possible} is subtle as the bulk too contains the gapless emergent photons. The corresponding dynamic structure factor is plotted in fig. \ref{fig_ybto_sf} with sizable spectral weight somewhat similar to that in experiments\cite{PhysRevLett.119.057203} at zero magnetic field.
\begin{figure}
\centering
\subfigure[Coupling constants ~~~~~~~~from Ref.~\onlinecite{PhysRevLett.119.057203}]{
\includegraphics[scale=0.225]{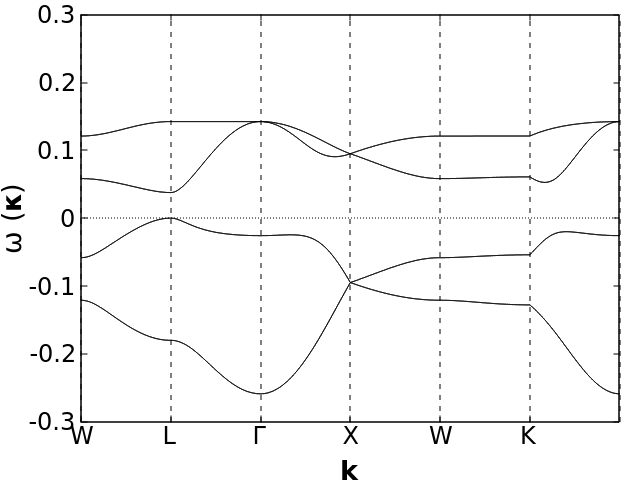}
}
\subfigure[Coupling constants ~~~~~~~~from Ref.~\onlinecite{ross2011quantum}]{
\includegraphics[scale=0.225]{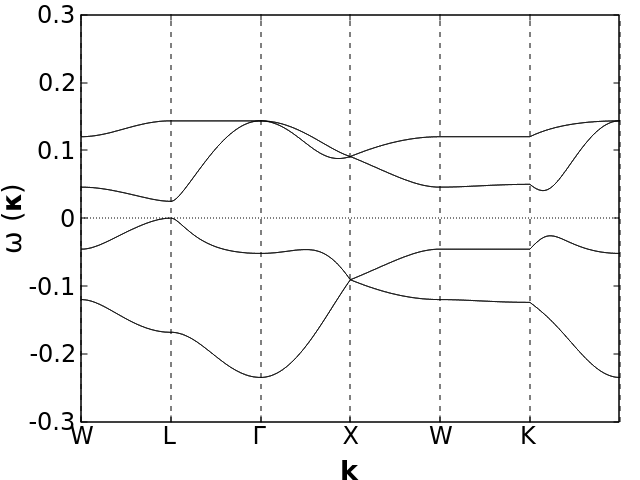}
}
\caption{Fermionic parton band structures for the experimental candidates with competing uniform U(1) QSL and magnetic order, the parameters are given in Table \ref{tab_coupl}. The independent mean field parameters are obtained by minimising $\tilde{E}_{MF}$ in \eqn{eqn:EMF_tot} and by setting the chemical potential self consistently satisfying the constraint in \eqn{eqn:mu_tot}. The dotted line in each figures denotes the chemical potential. In both the cases no magnetic order is present and the parton band structure has a finite excitation gap.}
\label{fig:bd_ybto}
\end{figure}

\begin{figure}
\centering
\includegraphics[scale=0.45]{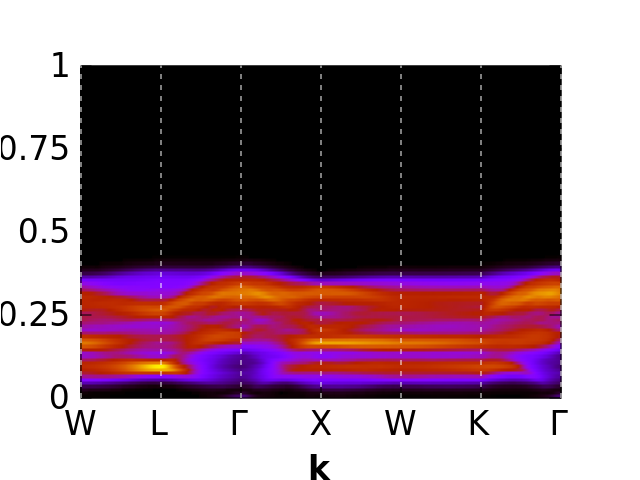}
\caption{The dynamical structure factor corresponding to the band structure in fig. \ref{fig:bd_ybto}(b) for Yb$_2$Ti$_2$O$_7$ within the pure QSL state obtained within a self-consistent parton mean field theory. The corresponding figure obtained from \ref{fig:bd_ybto}(a) (not shown) looks identical within our resolution.}
\label{fig_ybto_sf}
\end{figure}  

\paragraph*{{\rm Er$_2$Sn$_2$O$_7$ :}} Classical calculations places {\rm Er$_2$Sn$_2$O$_7$ :} near the phase boundary between the Palmer-Chalker state and the XY antiferromagnet. Our calculations reveal an interesting possibility of it being in a fractionalised Palmer-Chalker magnetic phase (\ref{fig_pd}). As reviewed in the introduction, experiments see Palmer-Chalker type magnetic order below $108$ mK\cite{PhysRevB.97.024415,PhysRevLett.119.187202} with a sizable {\it quasi-elastic} contribution in the to the low energy spectral weight just below the ordering temperature. Indeed such diffuse features can indeed arise from fractionalised partons which, according to the present calculations can coexist with the Palmer-Chalker state. In fig. \ref{fig:bd_m2}(a), we plot the parton band structure in the fractionalised Palmer-Chalker state (PC*) which is obtained within our self-consistent calculations for experimentally evaluated exchange couplings. The corresponding structure factor is plotted in fig. \ref{fig:bd_m2}(b) which shows flat band of diffused scattering similar to the broadened spectral weight seen in experiments.\cite{PhysRevLett.119.187202}

In addition to the above two, the neutron scattering data on powder of Er$_2$Pt$_2$O$_7$ also suggest strong suppression of ordering temperature to Palmer-Chalker state along with broad and anomalous diffused excitations at finite energies.\cite{PhysRevLett.119.187201} Present estimation of the exchange couplings also puts in in a regime where competition between the uniform $U(1)$ QSL and the Plamer-Chalker state is relevant (see fig. \ref{fig_pd}).

\begin{figure}
\centering
\subfigure[]{
\includegraphics[scale=0.225]{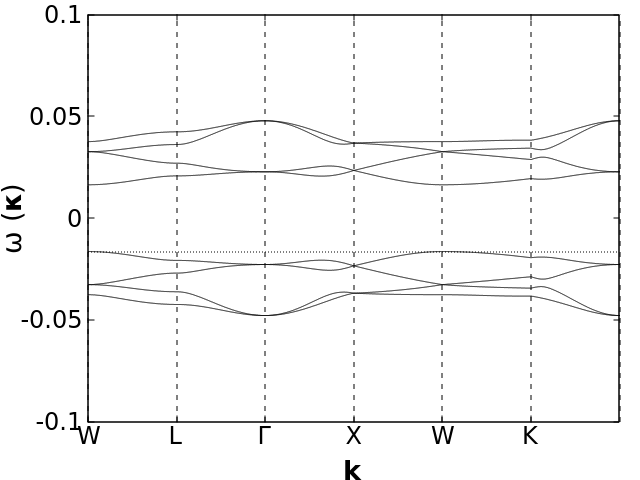}
}
\subfigure[]{
\includegraphics[width=0.48\linewidth,height=3.0cm]{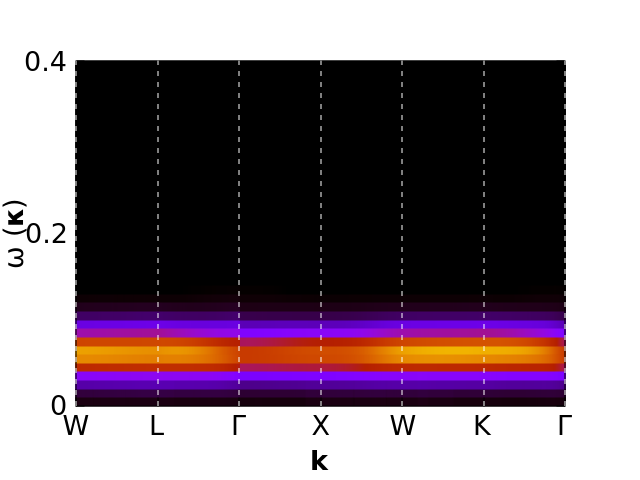}
}
\caption{Fermionic parton band structure(a) and dynamical structure factor for Er$_2$Sn$_2$O$_7$ (parameters given in Table \ref{tab_coupl}). The independent mean field parameters are obtained by minimising $\tilde{E}_{MF}$ in \eqn{eqn:EMF_tot} and by setting the chemical potential self consistently satisfying the constraint in \eqn{eqn:mu_tot}. Our self consistant mean field solution shows coexistance of magnetic order and U(1) QSL, with a finite excitation gap in the parton band structure. The dotted line in the band structure \ref{fig:bd_m2}$(a)$ denotes the chemical potential.
}
\label{fig:bd_m2}
\end{figure}

\section{Summary and outlook}
\mylabel{sec:conclusion}

In this work we have explored the unconventional magnetism in rare-earth pyrochlores starting with a uniform time reversal invariant U(1) QSL and exploring its instabilities to  magnetically ordered phases within a self-consistent fermionic parton mean field theory.  Our calculation suggests that in the uniform QSL, the fermionic parton band-structure can be gapped or gapless. In the former case the parton band-structure has a non-zero $Z_2$ topological invariant. This gapped phase is thus same as the time-reversal symmetry enriched U(1) QSL proposed in Ref. \onlinecite{PhysRevX.6.011034}-- the so called $(E_{fT}M_f)_\theta$ phase in their nomenclature. Such a phase has gapless surface states (that can lead to finite contribution to thermal conductivity at low temperatures) in addition to gapless photons in the bulk. 

Our studies regarding the magnetic instabilities of the above QSL to the ${\bf q}=0$ magnetic states obtained in the classical limit reveals a very rich phase diagram (fig. \ref{fig_pd}). Near the classical phase boundary between the magnetically ordered phases, the uniform $U(1)$ QSL becomes competitive and can open up regions of QSL as well as fractionalised magnetically ordered phases. The low energy diffused neutron scattering in materials such as Yb$_2$Ti$_2$O$_7$, Er$_2$Sn$_2$O$_7$ and Er$_2$Pt$_2$O$_7$ at zero/low magnetic field can arise due to proximity of the material to the phase boundary between magnetically ordered phase and the uniform U(1) QSL or due to fractionalised magnetically ordered phases. 

In addition to competing magnetic orders, the quantum spin-ice, the coulomb ferromagnet and the coexisting FM with monopole flux state has been proposed as an alternate starting point to understand the magnetic behaviour of Yb$_2$Ti$_2$O$_7$.  Our studies adds to this understanding by providing a different QSL as the starting point. Future numerical calculations such as variational Monte-Carlo methods may be able to identify which of these proposals is best suited to capture the relevant physics.
\section*{Acknowledgement}
We acknowledge Y. B. Kim, R. Moessner, P. McLarty, J. Rau, J. Roy, A. Agarwala and A. Nanda for fruitful discussions. S.B. acknowledges MPG for funding through the partner group on strongly correlated systems at ICTS and support of the SERB-DST (India) early career research grant (ECR/2017/000504). S. S. acknowledge funding through SERB-DST (India), Indo-US Science and Technology Forum, Indo-US post-doctoral fellowship. S. S. and K. D. acknowledges ICTS, TIFR where the project was initiated. We acknowledge {\it boson} and {\it zero} computing facilities at ICTS, TIFR.
\appendix
\section{Details of the lattice and the Hamiltonian}
\label{appen_ham_lattice}
\subsection{Unit cell and basis vectors of the pyrpchlore lattice}

The rare-earth cubic pyrochlores belong to the space group Fd$\bar{3}$m.\cite{gardner2010magnetic}  The local spin moments of the rare earth R$^{3+}$ sit on the vertices of a network of corner sharing tetrahedron. There are two types of tetrahedron as shown in Fig. \ref{fig_pyro} denoted as the {\it up} and {\it down} tetrahedron respectively. By choosing an up tetrahedron (without loss of generality) as the origin of the underlying Bravais lattice, one can describe the pyrochlore network as a FCC lattice with 4-site unit cell forming the four sublattices as shown in fig. \ref{fig_pyro}. The basis vectors describing the lattice are:
\begin{align}
\textbf{a}_1=\frac{1}{2}[110],~\textbf{a}_2=\frac{1}{2}[101],~\textbf{a}_3=\frac{1}{2}[011],
\end{align}
the four-point basis is then given by
\begin{align}
\textbf{R}_1&= \frac{\textbf{a}_1+\textbf{a}_2+\textbf{a}_3}{4}\\
\textbf{R}_2&= \frac{-\textbf{a}_1+\textbf{a}_2+\textbf{a}_3}{4}\\
\textbf{R}_3&= \frac{\textbf{a}_1-\textbf{a}_2+\textbf{a}_3}{4}\\
\textbf{R}_4&= \frac{\textbf{a}_1+\textbf{a}_2-\textbf{a}_3}{4}
\end{align}

\subsection{Spins in local and global coordinates}
Interplay of strong intra-orbital coulomb interactions, spin-orbit coupling and the local crystal field  provides a natural direction for the spin quantization axis which is different at different sub-lattice sites.\cite{gardner2010magnetic} With this {\it local quantization direction}, the spin at sub-lattice site $i$ is given by
\begin{align}
{\bf S}_{i}=S^x_i\hat{\bf x}_i+S^y_i\hat{\bf y}_i+S^z_i\hat{\bf t}_i,
\end{align}
where, 
\begin{align}
\hat{\bf t}_1=\frac{[111]}{\sqrt{3}}, \hat{\bf t}_2=\frac{[\bar{1}\bar{1}1]}{\sqrt{3}}, 
\hat{\bf t}_3=\frac{[\bar{1}1\bar{1}]1}{\sqrt{3}}, \hat{\bf t}_4=\frac{[1\bar{1}\bar{1}]}{\sqrt{3}}
\end{align}
define the local axis of quantization in the $S^z$ diagonal basis for the $4$ sublattice sites and
\begin{align}
\hat{\bf x}_1=\frac{[\bar{2}11]}{\sqrt{6}}, \hat{\bf x}_2=\frac{[2\bar{1}1]}{\sqrt{6}}, \hat{\bf x}_3=\frac{[21\bar{1}]}{\sqrt{6}}, \hat{\bf x}_4=\frac{[\bar{2}\bar{1}\bar{1}]}{\sqrt{6}}\nonumber\\
\hat{\bf y}_1=\frac{[0\bar{1}1]}{\sqrt{2}}, \hat{\bf y}_2=\frac{[011]}{\sqrt{2}}, \hat{\bf y}_3=\frac{[0\bar{1}\bar{1}]}{\sqrt{2}}, \hat{\bf y}_4=\frac{[01\bar{1}]}{\sqrt{2}}
\end{align}
are the local transverse directions. The global axes are defined as 
\begin{align}
\hat{\bf g}_1=[100],~~\hat{\bf g}_2=[010],~~
\hat{\bf g}_3=[001]
\label{global_axes}
\end{align}
in which the spins are given by:
\begin{align}
{\bf s}_{i}=s^x_i\hat{\bf g}_1+s^y_i\hat{\bf g}_2+s^z_i\hat{\bf g}_3
\end{align}
\subsection{Transformation matrices for the spin vectors from local to global coordinates}
We can transform the spin components in local basis to global basis by the following rotation matrices
\begin{equation}
s_i^\alpha=R_i^{\alpha \beta}S_i^\beta,
\label{R_local_global}
\end{equation}
where $i(=1,2,3,4)$ are the sublattice-site indices and $\{\alpha,\beta\}\in x,y,z$. The transformation matrices ($R_i^{\alpha \beta}$) are given by: 
\bea
R_1&=&\left[
\begin{array}{ccc}
 -\sqrt{\frac{2}{3}} & 0 & \frac{1}{\sqrt{3}} \\
 \frac{1}{\sqrt{6}} & \frac{-1}{\sqrt{2}} & \frac{1}{\sqrt{3}} \\
 \frac{1}{\sqrt{6}} & \frac{1}{\sqrt{2}} & \frac{1}{\sqrt{3}}
\end{array}
\right]\nonumber,
R_2= \left[
\begin{array}{ccc}
 \sqrt{\frac{2}{3}} & 0 & \frac{-1}{\sqrt{3}} \\
 \frac{-1}{\sqrt{6}} & \frac{1}{\sqrt{2}} & \frac{-1}{\sqrt{3}} \\
 \frac{1}{\sqrt{6}} & \frac{1}{\sqrt{2}} & \frac{1}{\sqrt{3}}
\end{array}
\right]\\ \nonumber
R_3&=& \left[
\begin{array}{ccc}
 \sqrt{\frac{2}{3}} & 0 & \frac{-1}{\sqrt{3}} \\
 \frac{1}{\sqrt{6}} & \frac{-1}{\sqrt{2}} & \frac{1}{\sqrt{3}} \\
 \frac{-1}{\sqrt{6}} & \frac{-1}{\sqrt{2}} & \frac{-1}{\sqrt{3}}
\end{array}
\right]\nonumber,
R_4=\left[
\begin{array}{ccc}
 -\sqrt{\frac{2}{3}} & 0 & \frac{1}{\sqrt{3}} \\
 \frac{-1}{\sqrt{6}} & \frac{1}{\sqrt{2}} & \frac{-1}{\sqrt{3}} \\
 \frac{-1}{\sqrt{6}} & \frac{-1}{\sqrt{2}} & \frac{-1}{\sqrt{3}}
\end{array}
\right].
\eea

\subsection{Details of the Hamiltonian (eqn. \ref{eqn:globalH}) in global basis}
\label{Jglobal}
Following Refs. \onlinecite{PhysRevB.75.212404, PhysRevB.95.094422,ross2011quantum} the coupling matrices for various bonds are given by (see fig. \ref{fig_pyro} for numbering of the bonds):
\bea
&\mathcal{J}_{12}=\left[\begin{array}{ccc}
J_1 & J_3 & -J_4\\
J_3 & J_1 & -J_4\\
J_4 & J_4 & J_2\\
\end{array}\right], ~~
\mathcal{J}_{13}=\left[\begin{array}{ccc}
J_1 & -J_4 & J_3\\
J_4 & J_2 & J_4\\
J_3 & -J_4 & J_1\\
\end{array}\right],\nonumber\\
\nonumber\\
&\mathcal{J}_{14}=\left[\begin{array}{ccc}
J_2 & J_4 & J_4\\
-J_4 & J_1 & J_3\\
-J_4 & J_3 & J_1\\
\end{array}\right],~~
\mathcal{J}_{23}=\left[\begin{array}{ccc}
J_2 & J_4 & -J_4\\
-J_4 & J_1 & -J_3\\
-J_4 & -J_3 & J_1\\
\end{array}\right],\nonumber\\
\nonumber\\
&\mathcal{J}_{24}=\left[\begin{array}{ccc}
J_1 & -J_4 & -J_3\\
J_4 & J_2 & -J_4\\
-J_3 & J_4 & J_1\\
\end{array}\right],~~
\mathcal{J}_{34}=\left[\begin{array}{ccc}
J_1 & -J_3 & -J_4\\
-J_3 & J_1 & J_4\\
J_4 &-J_4 & J_2\\
\end{array}\right]\nonumber.
\eea
\subsection*{Details of the Hamiltonian (eqn. \ref{eqn:localH}) in local basis}
The $\zeta$ matrix used in eqn. \ref{eqn:localH} is given by
\beq
\zeta=
\left(
 \begin{array}{cccc}
 0 & e^{-\frac{i \pi }{3}} & e^{\frac{i \pi }{3}} & -1 \\
 e^{-\frac{i \pi }{3}} & 0 & -1 & e^{\frac{i \pi }{3}} \\
 e^{\frac{i \pi }{3}} & -1 & 0 & e^{-\frac{i \pi }{3}} \\
 -1 & e^{\frac{i \pi }{3}} & e^{-\frac{i \pi }{3}} & 0 \\
\end{array}
\right).
\label{zeta}
\eeq 
The relations between the coupling constants in the global and local descriptions are\cite{ross2011quantum}
 \bea
 J_{zz}&=-\frac{1}{3}\left[2 J_1-J_2+2 (J_3+2 J_4)\right]\nonumber\\
 J_{\pm}&=\frac{1}{6} \left[2 J_1-J_2-J_3-2 J_4\right]\nonumber\\
 J_{\pm\pm}&=\frac{1}{6}\left[J_1+J_2-2 J_3+2 J_4\right]\nonumber\\
 J_{z\pm}&=\frac{1}{3 \sqrt{2}}\left[J_1+J_2+J_3-J_4\right]
 \label{eq_glob_loc}
 \eea

\subsection{Various coupling constants}
\label{appen_coupling}
Here we summarise the values of the experimentally measured exchange couplings for different materials of our interest in Table \ref{tab_coupl}.
\begin{table}[!htb]
\begin{tabular}{|c|c|c|c|c|c|}\hline
&\multicolumn{2}{c |}{Yb$_2$Ti$_2$O$_7$}&Er$_2$Ti$_2$O$_7$ & Er$_2$Sn$_2$O$_7$ &Er$_2$Pt$_2$O$_7$\\ \hline\hline
$J_1$ &-0.09 &-0.028&0.11&0.07&0.1$\pm 0.05$\\ \hline
$J_2$ &-0.22 &-0.326&-0.06&0.08&0.2$\pm 0.05$\\ \hline
$J_3$ &-0.29 &-0.272&-0.10&-0.11&-0.10$\pm0.03$\\ \hline
$J_4$ &-0.01 &0.049&-0.003&0.04&0\\ \hline\hline
$J_{zz}$ &0.17 &0.026&-0.025&0&$\sim$0.1\\ \hline
$J_{\pm}$ &0.05 &0.074&0.065&0.014&$\sim 0.017$\\ \hline
$J_{\pm\pm}$ &0.05 &0.048&0.042&0.074&$\sim 0.083$\\ \hline
$J_{z\pm}$ &-0.14 &-0.159&-0.009&0&$\sim0.047$\\ \hline
\end{tabular}
\caption{Experimentally measured coupling constants in meV for selected rare-earth pyrochlores. For Yb$_2$Ti$_2$O$_7$, the first column is from Ref. \onlinecite{ross2011quantum} whereas the second column is from Ref. \onlinecite{PhysRevLett.119.057203}.}
\label{tab_coupl}
\end{table}

The relation between the above coupling constants and there relation with the coefficients for the magnetic coupling constants in \eqn{eqn:H_MO} is given in Table \ref{a_coeff}.

\begin{table*}[!htb]
\begin{tabular}{  ccc }
\hline
     $a$~ & 
    $local$ ~ & 
    $global$~\\
\hline
\hline
$a_{A_2}$ & $3 J_{zz}$ & $-2 J_1 + J_2 -2( J_3 + 2 J_4) $ \vspace{3mm}
  \\
$a_E$ & $-6 J_{\pm}$ & $-2 J_1 + J_2 + J_3 + 2 J_4 $ \vspace{3mm}
  \\
$  a_{\text{T1A}'}$ &  $\frac{1}{6}(6 J_{\pm}+12 J_{\pm\pm}-3
   J_{zz})$\\&$ -\frac{1}{3} \sin (2 t) \left(2 \sqrt{2} J_{\pm}+4 \sqrt{2} J_{\pm\pm}-4 J_{z\pm}+\sqrt{2} J_{zz}\right)$\\ &$+ \frac{1}{6}\cos
   (2 t) \left(2 J_{\pm}+4 J_{\pm\pm}+16 \sqrt{2} J_{z\pm}+J_{zz}\right)$  & \begin{tabular}{@{}c@{}}$(2J_1+J_2)\cos^2(t) - (J_2+J_3-2 J_4)\sin^2(t)$\\$+\sqrt{2}J_3\sin(2t)$\end{tabular} \vspace{3mm}
\\
$  a_{\text{T1B}'}$ &  $\frac{1}{6}(6 J_{\pm}+12 J_{\pm\pm}-3
   J_{zz})$\\&$ +\frac{1}{3} \sin (2 t) \left(2 \sqrt{2} J_{\pm}+4 \sqrt{2} J_{\pm\pm}-4 J_{z\pm}+\sqrt{2} J_{zz}\right)$\\ &$- \frac{1}{6}\cos
   (2 t) \left(2 J_{\pm}+4 J_{\pm\pm}+16 \sqrt{2} J_{z\pm}+J_{zz}\right)$  & \begin{tabular}{@{}c@{}}$(2J_1+J_2)\sin^2(t) - (J_2+J_3-2 J_4)\cos^2(t)$\\$-\sqrt{2}J_3\sin(2t)$\end{tabular} \vspace{3mm}
\\
$a_{T_2}$ &$2 (J_{\pm}-2 J_{\pm\pm}) $ & $-J_2+J_3-2J_4$ \vspace{3mm}\\
\hline
\end{tabular}
\caption{Coefficients of the classical magnetic order parameter Hamiltonian in \eqn{eqn:H_MO}.}
\label{a_coeff}
\end{table*}

 \section{Transformation of parton operators from local to global basis}
\label{appen_loc_glob}
The global axes can be transformed to the local axes at each lattice site by the corresponding sub-lattice dependent three Euler rotations $R(\theta,\hat{\textbf{n}})$ (rotation matrix for an anti-clockwise rotation of $\theta$ about the axis $\hat{\textbf{n}}$), in Euler axis-angle representation. To find the transformation of the partons in the global basis to local basis, we use the spin-1/2 representation of these Euler rotations, $U(R(\theta,\hat{n}))=e^{-i \frac{\theta}{2}\vec{\sigma}.\hat{\textbf{n}}}$ on the two component parton fields.


The corresponding operator for transformation of partons from the local ($f$) to global ($F$) basis is defined as
\beq\mylabel{eqn:globaltolocal_spinon}
f_{i\sigma}=V_{i;\sigma\sigma'}F_{i\sigma'}.
\eeq  
Where $i(=1,2,3,4)$ is the sub-lattice index. For $i=1$ we show the operator $V_{1;\sigma\sigma'}$ as
\bea
V_{1}&=U^{\dagger}\left(R\left(\frac{\pi}{4},\hat{\bf g}_3\right)\right)U^{\dagger}\left(R\left(\cos^{-1}(\frac{1}{\sqrt{3}}),\frac{\hat{\bf g}_3 \times \hat{\bf t}_1)}{|\hat{\bf g}_3 \times \hat{\bf t}_1)|}\right)\right)\nonumber\\
&\hspace{4cm}U^{\dagger}\left(R\left(\frac{2\pi}{3},\hat{\bf t}_1\right)\right).
\eea
The form of $V_{i;\sigma\sigma'}$ for other sub-lattice indices can be worked out similarly. Using \eqn{eqn:MFTchannels} and \eqn{eqn:globaltolocal_spinon} we derive the relations between the singlet and triplet  particle-hole channels in local and global basis.

\section{Symmetry transformations}
\label{appen_symmetries}
The tetrahedral point-group of pyrochlore $\mathcal{T}_d$ consists 24 symmetry operations as $\{E, 8C_3, 3 C_2, 6 S_4, 6 \sigma_d\}$ where
 $C_3 \rightarrow \pm\frac{2\pi}{3}$ rotations around a local $[111]$ axis, $C_2 \rightarrow \pm\frac{\pi}{2}$ rotations around a local $[100]$ axis,$S_4 \rightarrow \pm\frac{\pi}{2}$ rotations around a local $[100]$ axis and then a reflection in the same $[100]$ plane,$\sigma_d \rightarrow$ reflection in  $[011]$ plane, $E \rightarrow$ identity.  $\mathcal{I}\in \{E,I\}$ (with $E$ being the identity and $I$ the inversion) denotes the symmetry between the up pointing and down pointing tetrahedron under inversion.
 
The transformation of the spin operators under thees symmetries can be worked out\cite{PhysRevB.96.125145} which forms the basis for understanding the transformation of the partons and hence the mean-field decoupling channels $\bar\chi_{ij}$ and $\bar{E}_{ij}^\alpha$.

\subsection{Symmetry transformation of the parton bilinears}
\label{symmetry_parton}
The transformations of the parton singlet and triplet operators under the above symmetry transformations give (in the local basis)

\bea
E_{12}^y &\rightarrow & \sqrt{3} E_{12}^x,~~~~\chi_{12}\rightarrow 0\\ \nonumber
E_{13}^y &\rightarrow &-\sqrt{3} E_{12}^x,~E_{13}^x\rightarrow -E_{12}^x,~E_{13}^z\rightarrow -E_{12}^z,~\chi_{13}\rightarrow 0\\ \nonumber
E_{14}^y &\rightarrow & 0,~E_{14}^x\rightarrow -2 E_{12}^x,~E_{14}^z\rightarrow  E_{12}^z,~\chi_{14}\rightarrow 0\\ \nonumber
E_{23}^y &\rightarrow & 0,~E_{23}^x\rightarrow -2 E_{12}^x,~E_{23}^z\rightarrow  E_{12}^z,~\chi_{23}\rightarrow 0\\ \nonumber
E_{23}^y &\rightarrow & 0,~E_{23}^x\rightarrow -2 E_{12}^x,~E_{23}^z\rightarrow  E_{12}^z,~\chi_{23}\rightarrow 0\\ \nonumber
E_{34}^y &\rightarrow & \sqrt{3}E_{34}^x,~E_{34}^x\rightarrow  E_{12}^x,~E_{34}^z\rightarrow  E_{12}^z,~\chi_{34}\rightarrow 0
\label{table:symmetry_local}
\eea

These relations shows that there are only two independent singlet/triplet operators which can have non-zero mean field value. Under inversion symmetry, an up-pointing tetrahedron are mapped to a down pointing tetrahedron. For the action of inversion on singlet and triplet operators, consider the tetrahedron in fig \ref{fig_pyro}.
On inversion about site $1$;  $E^\alpha_{1,j}\rightarrow E^\alpha_{1,j'} \text{and} \, \chi_{1,j}\rightarrow \chi_{1,j'}$; where $j'$ 
is the site that $j$ site maps to under inversion.

\subsection{G-matrices in global basis}\label{gmatrix}
 In local basis $g_{\mu\nu}^a$ is $g-tensor$ given by the diagonal matrix 
 \begin{align}
 {\rm Diag}\{g_{xy},g_{xy},g_{zz}\}
 \label{appen_g_local}
 \end{align}
 for all $a$.
The $G$ matrices in global basis, obtained by doing the basis transformations for various sublattice sites as $G_{\alpha \beta}^a=R_{\alpha \gamma}^a g_{\gamma \delta}^a (R_{\delta \beta}^a)^T$, are as given below, with $g_1=\frac{2 g_{\text{xy}}}{3}+\frac{g_{\text{zz}}}{3},\, g_2=-\frac{g_{\text{xy}}}{3}+\frac{g_{\text{zz}}}{3}$.
\bea
G_1&=&\left[
\begin{array}{ccc}
 g_1 & g_2 & g_2 \\
 g_2 & g_1 & g_2 \\
 g_2 & g_2 & g_1
\end{array}
\right]\nonumber,
G_2=\left[
\begin{array}{ccc}
 g_1 & g_2 & -g_2 \\
 g_2 & g_1 & -g_2 \\
 -g_2 & -g_2 & g_1
\end{array}
\right]\\ \nonumber
G_3&=&\left[
\begin{array}{ccc}
 g_1 & -g_2 & g_2 \\
 -g_2 & g_1 & -g_2 \\
 g_2 & -g_2 & g_1
\end{array}
\right]\nonumber, 
G_4=\left[
\begin{array}{ccc}
 g_1 & -g_2 & -g_2 \\
 -g_2 & g_1 & g_2 \\
 -g_2 & g_2 & g_1
\end{array}
\right]\nonumber.
\eea
\section{Details of the parton mean field theory}
\subsection{Details for \eqn{eqn:HMF} and \eqn{HMF_MO_terms}}
\label{appen_MFT_details}
Here we provide the expressions of the terms of the Hamiltonian $H^{ab}_{\sigma_1 \sigma_2}$ in \eqn{eqn:HMF}. In the equations below $\bf{R}_{ij}\equiv (\bf{R}_i-\bf{R}_j)/2$. 
\bea
H^{12}_{\uparrow\uparrow}&=& F\cos
   (\textbf{k}.\textbf{R}_{12})=-H^{12}_{\downarrow\downarrow}\\ \nonumber
   H^{12}_{\uparrow\downarrow}&=& G \cos
   (\textbf{k}.\textbf{R}_{12})=-(H^{12}_{\downarrow\uparrow})^*\\ \nonumber
   H^{13}_{\uparrow\uparrow}&=& -F\cos
   (\textbf{k}.\textbf{R}_{13})=-H^{13}_{\downarrow\downarrow}\\ \nonumber
   H^{13}_{\uparrow\downarrow}&=& G^* \cos
   (\textbf{k}.\textbf{R}_{13})=-(H^{13}_{\downarrow\uparrow})^*\\ \nonumber
   H^{14}_{\uparrow\uparrow}&=& F\cos
   (\textbf{k}.\textbf{R}_{14})=-H^{14}_{\downarrow\downarrow}\\ \nonumber
   H^{14}_{\uparrow\downarrow}&=& H \cos
   (\textbf{k}.\textbf{R}_{14})=H^{14}_{\downarrow\uparrow}\\ 
\nonumber
   H^{23}_{\uparrow\uparrow}&=& F\cos
   (\textbf{k}.\textbf{R}_{23})=-H^{23}_{\downarrow\downarrow}\\ \nonumber
   H^{23}_{\uparrow\downarrow}&=& H \cos
   (\textbf{k}.\textbf{R}_{23})=H^{23}_{\downarrow\uparrow}\\ 
\nonumber
H^{34}_{\uparrow\uparrow}&=& F\cos
   (\textbf{k}.\textbf{R}_{34})=-H^{34}_{\downarrow\downarrow}\\ \nonumber
   H^{34}_{\uparrow\downarrow}&=& G \cos
   (\textbf{k}.\textbf{R}_{34})=-(H^{34}_{\downarrow\uparrow})^*\\ \nonumber
\eea
where
\bea
F&=&\frac{i}{4}(8 E_x J_{z\pm}+E_z (4 J_{\pm}+J_{zz}))\\ \nonumber
G&=&\frac{i+\sqrt{3}}{4} (4 E_xJ_{\pm\pm}+2 E_z J_{z\pm}-E_x J_{zz})\\ \nonumber
H&=&-\frac{i}{2}(4 E_x J_{\pm\pm}+2 E_z J_{z\pm}-E_x J_{zz})
\eea
To obtain \eqn{HMF_MO_terms}, we need to add the magnetic decoupling channels to the pure QSL in \eqn{eqn:HMF}. Keeping in mind \eqn{eq_hybrid}, the additional matrix elements for the augmented parton mean field theory is given by 

\bea
H^{11}_{\uparrow \uparrow}&=&-H^{11}_{\downarrow \downarrow}=-H^{33}_{\uparrow \uparrow}=H^{33}_{\downarrow \downarrow}=4\alpha I\\ \nonumber
H^{11}_{\uparrow \downarrow}&=&4\alpha J,~~H^{33}_{\uparrow \downarrow}=4\alpha K\\ \nonumber
H^{22}_{s_1 s_2}&=&H^{11}_{s_1 s_2},~~H^{44}_{s_1 s_2}=H^{33}_{s_1 s_2}
\eea
where $\lbrace s_1,s_2 \rbrace\equiv \lbrace  \uparrow,\downarrow \rbrace $ and
\bea
I&=&\bar{m}_{\text{T}_{1A'}}~a_{\text{T1A}'} \left(\frac{\sin (t)}{4\sqrt{6}}+\frac{\cos (t)}{8 \sqrt{3}}\right)\\ \nonumber
J&=& \left(\frac{\sqrt{3}}{16}-\frac{i}{16}\right) \bar{m}_{\text{E}} a_E +
\left(\frac{\sqrt{3}}{16}+\frac{i}{16}\right) \bar{m}_{\text{T2}}
   ~a_{\text{T2}} \nonumber\\
    &+&
   \bar{m}_{\text{T}_{1A'}} ~a_{\text{T1A}'} \biggl(\left(\frac{1}{4
   \sqrt{6}}-\frac{i}{4 \sqrt{2}}\right) \cos (t)\nonumber\\ \nonumber
   &-&\left(\frac{1}{8
   \sqrt{3}}-\frac{i}{8}\right) \sin (t)\biggr)\\ \nonumber
K&=& \left(\frac{\sqrt{3}}{16}-\frac{i}{16}\right) \bar{m}_{\text{E}} a_E -
\left(\frac{\sqrt{3}}{16}+\frac{i}{16}\right) \bar{m}_{\text{T2}}
   ~a_{\text{T2}} \nonumber\\
    &-&
   \bar{m}_{\text{T}_{1A'}} ~a_{\text{T1A}'} \biggl(\left(\frac{1}{4
   \sqrt{6}}-\frac{i}{4 \sqrt{2}}\right) \cos (t)\nonumber\\ \nonumber
   &-&\left(\frac{1}{8
   \sqrt{3}}-\frac{i}{8}\right) \sin (t)\biggr)
\eea

\subsection{Minimization}
\label{appen_mft_min}
In order to search for the mean-field ground state by minimizing $E_{MF}$(say \eqn{eqn:EMF}) we first scan the parametric $\bar{E}_x-\bar{E}_z$ plane for bounded mean-field solutions. The condition for a bounded minimum in $E_{MF}$ is given in TABLE \ref{stability}, where $\mathbb{Q}$ is the stability matrix given in \eqn{eqn:Q}; obtained by expressing the quadratic part of $E_{MF}$(\eqn{eqn:EMF}) as $\begin{pmatrix} \bar{E}_z\bar{E}_x \end{pmatrix} \mathbb{Q} \begin{pmatrix} \bar{E}_z\\\bar{E}_x \end{pmatrix}$.

\begin{table}[!htb]
\begin{tabular}{ccc}
\\
\hline
$Det(\mathbb{Q})>$ 0 & $Tr(\mathbb{Q})\leq 0$ & \begin{tabular}{@{}c@{}}no bounded mean field \\ ground state\end{tabular}\\
$Det(\mathbb{Q})>$ 0 & $Tr(\mathbb{Q})> 0$ & \begin{tabular}{@{}c@{}}mean field ground state \\ in $\bar{E}_x-\bar{E}_z$ plane \end{tabular}\\
$Det(\mathbb{Q})\leq $ 0 & ~~$Tr(\mathbb{Q})\leq or > 0$ & \begin{tabular}{@{}c@{}}mean field ground state in \\ $\bar{E}_x(\bar{E}_z)$ line \end{tabular}\\
\hline
\end{tabular}
\caption{Conditions for a bounded minima in $E_{MF}$}
\label{stability}
\end{table}

\beq\mylabel{eqn:Q}
\mathbb{Q}=\begin{pmatrix} 2(J_{zz}+4J_{\pm})&16 J_{z\pm}\\16 J_{z\pm}&8(-J_{zz} +4J_{\pm\pm}) \end{pmatrix}
\eeq

For a particular point in the phase space(given by $\{J_1,J_2,J_3,J4\}$ or $\{J_{zz},J_{\pm},J_{\pm\pm},J_{z\pm}\}$) if the conditions in TABLE \ref{stability} suggests that a bounded mean field minimum of $E_{MF}$ is located in  $\bar{E}_x-\bar{E}_z$ plane, we search for the minimum by an adaptive grid search method, where we first split the 2-dimensional parametric search space in an uniform grid and locate all the minima; followed by which we search in a small window in the parametric space around each of the minimum (obtained in the first step) with a finer grid. If the conditions in TABLE \ref{stability} suggests a bounded mean field solution is located along a line $\bar{E}_x(\bar{E}_z)$, we first identify the the unstable direction in phase space by noting that one of the eigenvalues of $\mathbb{Q}$ is $\leq$ 0. We set the corresponding eigenvector of $\mathbb{Q}$ to zero to find the bounded direction in phase space and then we apply the same adaptive grid search along that direction. At each step of the grid search we set the chemical potential $\mu$ by solving \eqn{eqn:mu} using standard bi-section method. 
\bibliography{biblio}
\end{document}